\documentclass[printer]{aa}

\usepackage{amsmath}
\usepackage[varg]{txfonts}

\begin{document}
\title{Investigating prominence turbulence with Hinode SOT Dopplergrams}

\author{A. Hillier\inst{1}, T. Matsumoto\inst{2} and K. Ichimoto\inst{3}}

\institute{CEMPS, University of Exeter, Exeter EX4 4QF U.K.\\ \email{a.s.hillier@exeter.ac.uk}
\and
Division of Physics and Astronomy, Graduate School of Science, Kyoto University
\and
Kwasan and Hida Observatories, Kyoto University}

\titlerunning{Using Hinode SOT to investigate prominence turbulence}
\authorrunning{Hillier et al.}

\abstract
{Quiescent prominences host a diverse range of flows, including Rayleigh-Taylor instability driven upflows and impulsive downflows, and so it is no surprise that turbulent motions also exist.
As prominences are believed to have a mean horizontal guide field, investigating any turbulence they host could shed light on the nature of magnetohydrodynamic (MHD) turbulence in a wide range of astrophysical systems.
In this paper we have investigated the nature of the turbulent prominence motions using structure function analysis on the velocity increments estimated from H$\alpha$ Dopplergrams constructed with observational data from Hinode Solar Optical Telescope (SOT).
The probability density function of the velocity increments shows that as we look at increasingly small spatial separations the distribution displays greater departure from a reference Gaussian distribution, hinting at intermittency in the velocity field.
Analysis of the even order structure functions for both the horizontal and vertical separations showed the existence of two distinct regions displaying {different exponents of the power law} with {the break in the power law} at approximately $2000$\,km.
We hypothesise this {to be a result} of internal turbulence excited in the prominence by the dynamic flows of the system found at this spatial scale.
We found that the scaling exponents of {the $p^{\rm th}$ order} structure functions for these two regions generally followed the $p/2$ (smaller scales) and $p/4$ (larger scales) {laws} that are the same as those predicted for weak MHD turbulence and Kraichnan-Iroshnikov turbulence respectively.
{However, the existence of the $p/4$ scaling at larger scales than the $p/2$ scaling is inconsistent with the increasing nonlinearity expected in MHD turbulence.}
We also found that as we went to higher order structure functions, the dependence of the scaling exponent on the order $p$ is nonlinear implying that intermittency may be playing an important role in the turbulent cascade.
Estimating the heating from the turbulent energy dissipation showed that this turbulence would be very inefficient at heating the prominence plasma, but that the mass diffusion through turbulence driven reconnection was of the order of $10^{10}$\,cm$^2$\,s$^{-1}$. This is of similar order to that of the expected value of the ambipolar diffusion and a few orders of magnitude greater than Ohmic diffusion for a quiescent prominence.}

   \maketitle

\section{Introduction}

Prominences, cool dense clouds of partially ionised plasma supported above the solar surface by the Lorentz force, have long been known to host a wide range of dynamic motions including downflows \citep{ENG1981, KUBO1986} and {vortices} \citep{LZ1984}.
The launch of the Hinode satellite \citep{KOS2007} with the Solar Optical Telescope \citep[SOT; ][]{TSU2008}, shed new light onto the dynamics of prominences.
These new Hinode observations have developed out understanding of nonlinear prominence downflows \citep[e.g.][]{Chae2010, HILL2012b}, shown that prominences are full of a broad spectrum of magnetohydrodynamic (MHD) waves and oscillations \citep[e.g.][]{SCHM2010, SCHM2013, HILL2013} and the existence instabilities, for example the magnetic Rayleigh-Taylor instability \citep[e.g.][]{BERG2008, BERG2010, BERG2011, HILL2011, HILL2012}.

Considering their dynamic nature and the large Reynolds numbers it is no surprise that prominences also display the characteristics of a turbulent medium.
\citet{LEO2012} investigated the correlations in the Ca II H intensity in a quiescent prominence using Hinode SOT and these investigations revealed that there exists power laws in the wavenumber dependence of the power spectral density, with a break in the power law found at {approximately }$1000$\,km and signs of the multi-fractality of the prominences light curves.
{\citet{FREED2016} investigated the plane-of-sky velocity, obtained through feature tracking, determining the power spectra and found indices of the power law fit to the power spectra in the range $-1$ to $-1.6$.}

Turbulence itself is an area of fluid dynamics research of great interest.
The general nature of incompressible turbulence in a homogeneous, isotropic, and statistically steady system was first described by \citet{KOL1941}.
This process describes how energy injected at large scales cascades through progressively smaller scales until it reaches the dissipation scale of the system.
For Kolmogorov turbulence, simple dimensional analysis shows that the $p^{\rm th}$ order structure function of the velocity increments (as defined in Equation \ref{kolo}) in the inertial range follow the relation:
\begin{equation}\label{kolo}
<[v(\mathbf{x}+\mathbf{r})-v(\mathbf{x})]^p>=<\delta_rv^p> =C_p \epsilon^{p/3} r^{p/3},
\end{equation}
where $r$ is the distance over which the structure function is being calculated, $\epsilon$ is the energy dissipation rate and $C_p$ is a constant associated with the {$p^{\rm th}$ order structure function}.
The homogeneity, isotropy and statistical steadiness assumed by Kolmogorov are considered to hold for small-scale motions of turbulence even if the turbulence in the large scale is inhomogeneous, anisotropic and time-dependent. 
This is because memory of the large-scale anisotropy, for example, is likely to be lost during the energy cascade process which is roughly conceived as the process of successive splitting of large eddies to smaller ones. 
As a result, small eddies are considered to reach a universal state which is the one hypothesised by Kolmogorov, see for example, \citet{Davidson}.

Kolmogorov turbulence, however, does not deal with the influence of the magnetic field, something of great importance for many astrophysical systems.
\citet{KRAI1965} \& \citet{IROS1964} extended Kolmogorov turbulence to include the Alfv\'{e}n velocity ($V_{\rm A}$) in an isotropic scaling (hereafter K-I turbulence).
However, the inclusion of the Alfv\'{e}n velocity means that dimensional arguments can no longer produce a unique scaling.
Through arguments based on the number of wave interactions necessary to deform a wave packet, the K-I turbulence scaling is determined to be \citep[e.g.][]{KRAI1965}: 
\begin{equation}\label{KIspec}
<\delta_rv^p>=C'_p V_{\rm A}^{p/4} \epsilon^{p/4} r^{p/4}.
\end{equation}
This scaling is isotropic, meaning that there is no difference between the directions parallel and perpendicular to the magnetic field.
However, it can be expected that the presence of a strong mean magnetic field, as is likely to be the case for prominences, turbulence will not be isotropic.
This thought led to the development of anisotropic scalings in MHD turbulence.

For anisotropic MHD turbulence for a mean field magnetic field of $B_0$, two cases can be envisioned: The case of weak MHD turbulence where the perturbations to the magnetic field $b$ satisfy the condition $b \ll B_0$, and strong MHD turbulence where $b \sim B_0$.
In the case of weak MHD turbulence it has been shown that the turbulent cascade is dominated by the cascade {perpendicular} to the magnetic field, because the timescale for the deformation of the Alfv\'{e}n waves in the direction along the magnetic field is taken to be much longer than that perpendicular to the direction of the magnetic field \citep{NG1996}.
This gives the relation of the perpendicular velocity perturbations of \citep[e.g.][]{SCH2007}:
\begin{equation}\label{weak_MHD}
<\delta_{r\perp}v^p>=C''_p ( \epsilon V_{\rm A}/r_{\parallel 0})^{p/4} r_{\perp}^{p/2},
\end{equation}
where $\delta_{r\perp}v$ is the velocity increment calculated in the direction perpendicular to the magnetic field, $r_{\parallel}$ is the wavelength of the Alfv\'{e}n wave along the magnetic field (here we use $r_{\parallel 0}$ as there is no cascade along the magnetic field and as such the wavelength does not change throughout the turbulent cascade) and $r_{\perp}$ is the spatial separation perpendicular to the magnetic field at which the velocity increment is being calculated.

For weak MHD turbulence, as the cascade continues the characteristic timescale of the nonlinearity of the turbulence increases with respect to the frequency of the Alfv\'{e}n wave.
The nonlinearity of the turbulent fluctuations can be measured by the nonlinearity parameter $\chi_r$ given by:
\begin{equation}\label{nonlin}
\chi_r=\frac{\langle|\delta_{r \perp} v|\rangle \lambda}{V_{\rm A} r_{\perp}},
\end{equation} 
{where $\lambda$ is the wavelength of the {Alfv\'{e}n} wave and $\delta_{r\perp} v$ is now the characteristic velocity associated with $r_\perp$. Here we assume that the {Alfv\'{e}n} wave sits in a large scale characterised with the single wavelength $\lambda$ and the single velocity $V_{\rm A}$. The parameter $\chi_r$ can be interpreted as a ratio of the {Alfv\'{e}n} time scale over the nonlinear time scale \citep[see also][]{GOSR1995, GALT2005}.}
In the case of weak MHD turbulence $\chi_r \propto r_{\perp}^{-1/2}$, meaning that the nonlinearity of the turbulence increases as the scales across the magnetic field become smaller.
Eventually, given a sufficiently large inertial range, {there will exist an $r_{\perp}$ such that the value of $\chi_r$ becomes unity} and the turbulence transitions from weak turbulence to strong turbulence.

Once the turbulence transitions from weak to strong, the timescale for deformation of an Alfv\'{e}n wave is sufficiently short that the cascade is no longer solely resulting from fluctuations perpendicular to the magnetic field, though still dominated by the cascade in that direction.
Based on the concept of critical balance that at all scales of the strong MHD turbulence cascade $B_0/L=b/\delta${, where $L$ is the wavelength along the magnetic field of the Alfv\'en wave and $\delta$ is the amplitude of the displacement}, \citet{GOSR1995} found that the spectrum is anisotropic $k_{\parallel}\sim k_{\perp}^{2/3}L^{-1/3}$ and that the energy cascade is given as $E(k)\sim k_{\perp}^{-5/3}$, that is $<\delta_{r\perp}v^p>\sim r_{\perp}^{p/3}$.
However, there is still some controversy relating to the scaling of MHD turbulence.
\citet{BOLD2005} presented a new model based on the concept of dynamic alignment of vortices where $<\delta_{r\perp}v^p>\sim r_{\perp}^{p/4}$.

{Structure functions have been used as an important tool in determining the characteristics of turbulent and intermittent phenomena in the solar atmosphere and in the solar wind.
\citet{ABRA2002} and \citet{ABRA2002b} used structure functions to investigate the line-of-sight component of the magnetic field in active regions using SoHO/MDI and Huairou Solar Observing Station, China.
They found a departure from the linear scaling with order $p$ as a result of intermittency from the spatial fluctuations of the local dissipation rate \citep[e.g.][]{FRIS1996}.
The intermittency was found to increase with flaring activity in the region.
\citet{ABRA2010} extended these studies to include Hinode SOT/SP and Big Bear Solar Observatory data measuring the high-order structure functions for 214 solar active regions. They found that {a modified flatness function, defined in their paper as the }ratio of the sixth-order structure function to the cube of the second, below scales of $10$\,Mm was correlated with flare productivity.
\citet{BUCH2006} applied structure function analysis, amongst other analysis techniques, to SoHO/SUMER data, finding intermittency of the velocity field observed in ultraviolet light, with the results supporting the existence of small-scales created in the solar transition region by turbulence.
Application of structure function analysis to measurements of the velocity and magnetic fields in the solar wind also show the departures from the linear scaling with $p$ \citep{BRCA2013}.

The presence of a magnetic field, as well as the gravity driven flows, make prominences an interesting environment to investigate turbulence in an astrophysical system.
In this paper we investigate the velocity fluctuations of a prominence based on Dopplergrams created from Hinode SOT observations. 
We study the velocity increments, investigating how their distribution changes over different lengthscales, and determining the power laws associated with their structure functions to attempt to connect these velocity fluctuations with turbulence theory.
finally we estimate the energy deposition rate associated with this turbulence.

\section{Estimating the line-of-sight velocity of the prominence from Hinode SOT observations}

The target of the observations is a quiescent prominence seen on the NW solar limb (41$^{\circ}$N 84$^{\circ}$W) on 2008 September 29.
Using Hinode SOT this prominence was observed between 10:00UT and 14:00UT in H$\alpha$ $\pm 208$\,m{\AA} using the narrowband filter, as well as Ca II H in the broadband filter.
{In this observation sequence, the images in each line were taken at a cadence of 30\,s with a time separation between the images in the two wings of H$\alpha$ of 10\,s.}
In this paper we focus on the H$\alpha$ data.
Figure \ref{prom} panels (a) and (b) show the intensity in H$\alpha$ $\pm 208$\,m\AA.

\begin{figure*}[ht]
\centering
\includegraphics[width=5.5cm]{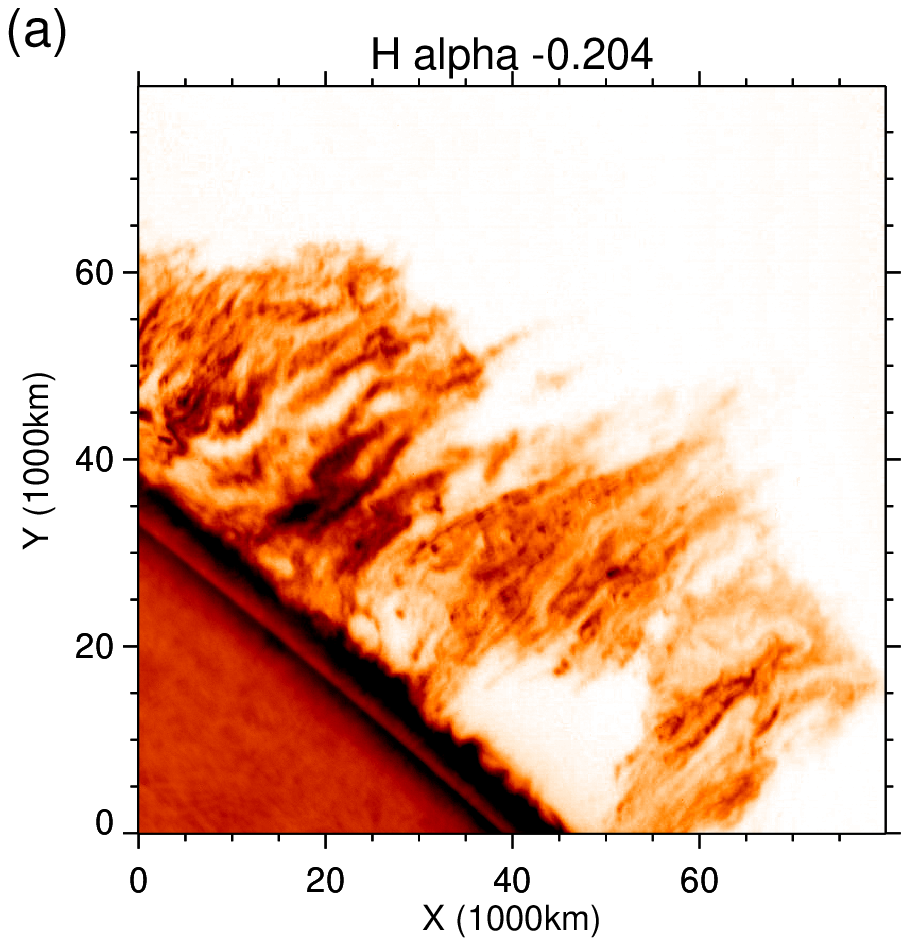}
\includegraphics[width=5.5cm]{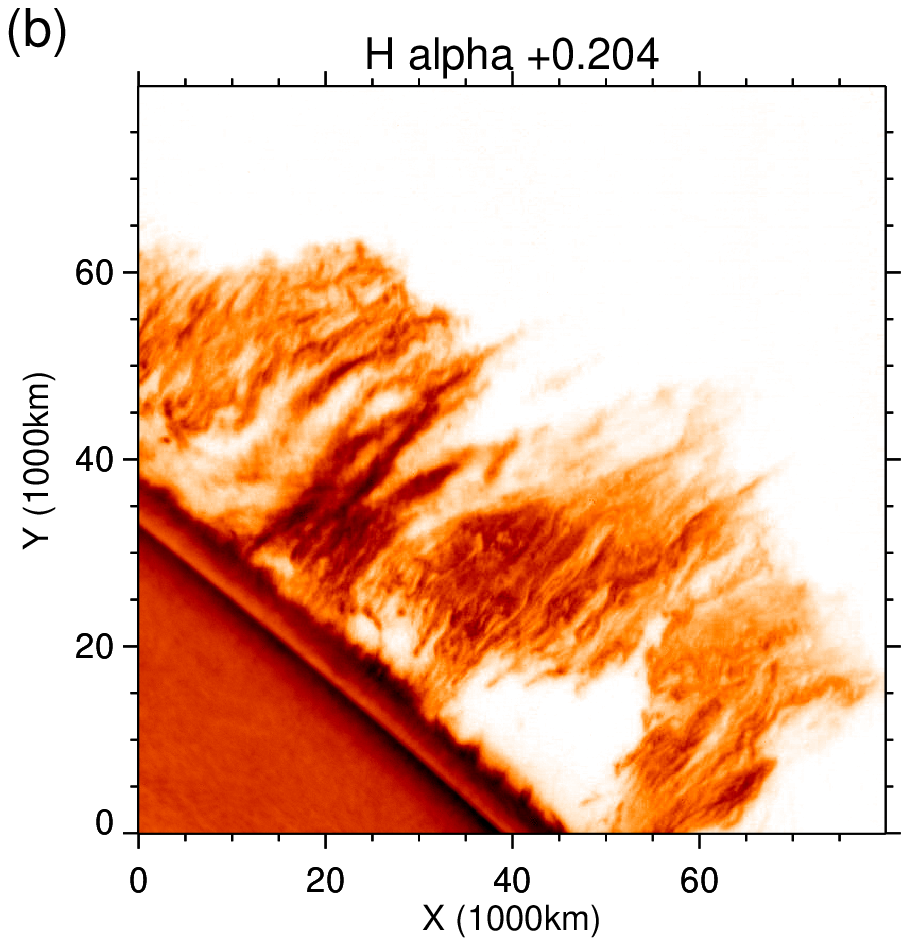}
\includegraphics[width=5.5cm]{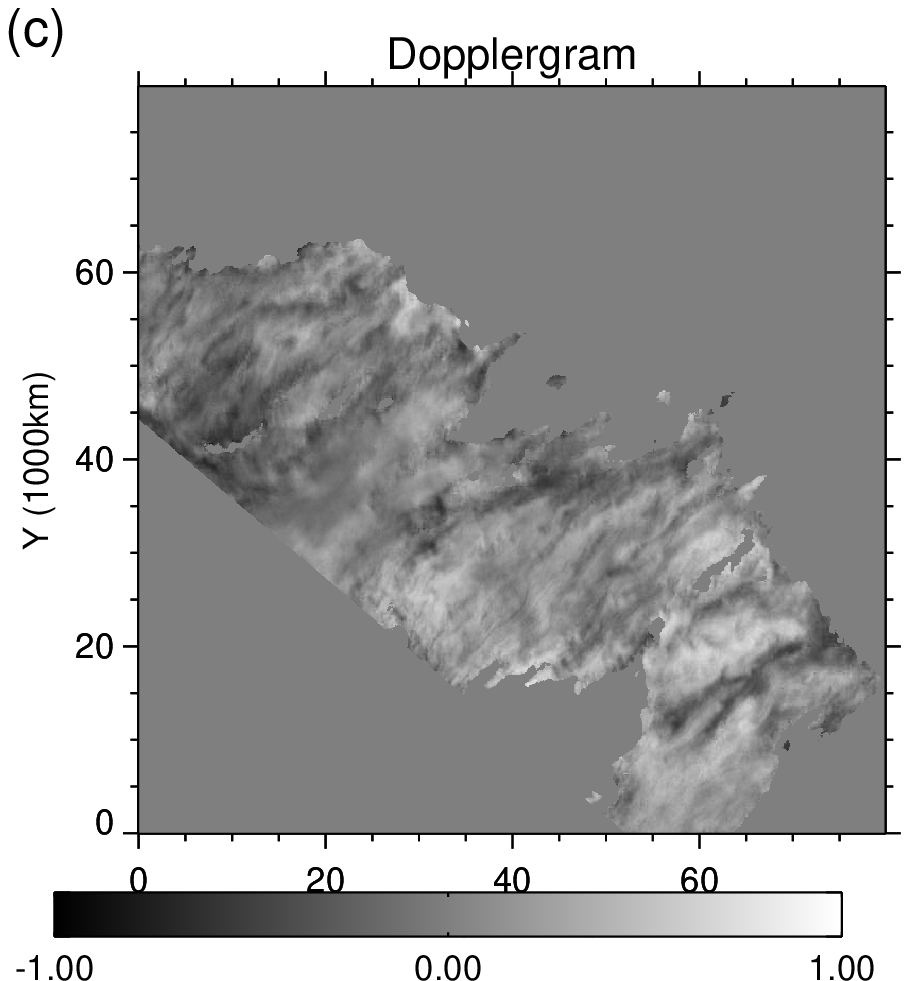}
\caption{(a) intensity in the H$\alpha+ 208$\,m{\AA} wing, (b) the intensity in the H$\alpha- 208$\,m{\AA} wing and (c) the Dopplergram at 2008-09-29 10:54:56 UT.}
\label{prom}
\end{figure*}

The H$\alpha$ Dopplergrams were created by subtracting line centre $+208$\,m{\AA} images from the nearest line centre $-208$\,m{\AA} images in time, then normalising by the sum of these images.
The equation to calculate this is:
\begin{equation}\label{dopgram_eqn}
I_{\rm D}=\frac{I_+-I_-}{I_++I_-}.
\end{equation}
The offset of $208$\,m{\AA} from line centre is equivalent to a Doppler velocity of $9.5$\,km\,{s}$^{-1}$.
The calculated Dopplergram is shown in panel (c) of Fig. \ref{prom}.
Here we would like to note that there is a component of {stray} light in the H$\alpha$ data that {if not removed} may have resulted in {significant changes in} the Dopplergram value as calculated using Equation \ref{dopgram_eqn}.
We present a detailed explanation in Appendix \ref{Appen1} regarding our techniques to process the data to remove {stray} light from the image data before calculating the Dopplergrams.

It is now necessary to connect the Dopplergram to a velocity, for this we use the cloud model \citep{BECK1964}.
From the cloud model, we can model the intensity as a function of wavelength ($I(\lambda)$) of the prominence as:
\begin{align}
I(\lambda)=S_{\lambda}\left[1-\exp(-\tau_{\lambda}) \right],\label{cloud}\\
\tau_{\lambda}=\tau_0 \exp\left[ -\frac{(\lambda-\lambda_{\rm D})^2}{\sigma_{\lambda}^2}  \right], \nonumber
\end{align}
where $S_{\lambda}$ is the source function, $\lambda$ is the wavelength, $\tau_0$ is the optical depth {at line centre giving $\tau_{\lambda}$ as the optical depth at wavelength $\lambda$}, {$\lambda_{\rm D}$ is the Dopplershifted position of line centre and $\sigma_{\lambda}$ is the line width.
Performing a Taylor expansion of $\exp(-\tau_{\lambda})$ up to the first term in $\tau_{\lambda}$, Equation \ref{cloud} becomes:
\begin{align}\label{cloud_new}
I(\lambda)=S_{\lambda}\tau_0 \exp\left[ -\frac{(\lambda'-\Delta \lambda)^2}{\sigma_{\lambda}^2}  \right],
\end{align}
where $\lambda'=\lambda-\lambda_0$ with $\lambda_0$ as the wavelength of the at rest line centre, and $\Delta \lambda$ is the shift in the position of the line from the rest wavelength.
{Here we note that $\lambda - \lambda_D = \lambda' - \Delta \lambda$.}
Physically speaking, taking the Taylor expansion is equivalent to assuming that the departure from the line profile of an optically thin plasma in the observed wavelengths (in this case in the red and blue wings centred on Dopplershifts of $\pm9.5$\,km\,s$^{-1}$) is small.
This assumption will lead to small errors in the estimation of the line shift, but will not result in a change of the sign nor of the relative magnitude of the line shift.
The ultimate result is the line profile being approximated by a Gaussian distribution.}

Based on this assumption that the line profile follows a Gaussian distribution, the value given by the Dopplergram can be related to a wavelength shift by the following formula:
\begin{equation}\label{hardeqn}
I_{\rm D}=\frac{\exp\left[-\frac{(\lambda_{\rm SOT}-\Delta \lambda)^2}{\sigma_{\lambda}^2} \right]-\exp\left[-\frac{(-\lambda_{\rm SOT}-\Delta \lambda)^2}{\sigma_{\lambda}^2} \right]}{\exp\left[-\frac{(\lambda_{\rm SOT}-\Delta \lambda)^2}{\sigma_{\lambda}^2} \right]+\exp\left[-\frac{(-\lambda_{\rm SOT}-\Delta \lambda)^2}{\sigma_{\lambda}^2} \right]},
\end{equation}
{where $I_{\rm D}$ is the value of the Dopplergram, $\lambda_{\rm SOT}$ is the wavelength offset from the rest line-centre of the Hinode SOT observations and $\sigma_{\lambda}$ is the width of the Gaussian distribution.
Equation \ref{hardeqn} can be simplified to:
\begin{equation}\label{vel_trans}
I_{\rm D}=\tanh \left(2\frac{ \lambda_{\rm SOT}\Delta \lambda}{\sigma_{\lambda}^2} \right).
\end{equation}
Here we can see} that, under the assumptions applied, the Dopplergram value is given as a function of the observed line position (i.e. $+208$\,m\AA), the Dopplershift of the line and the line width.

From Equation \ref{vel_trans} we are interested in obtaining $\Delta\lambda$ and we already know $\lambda_{\rm SOT}$, but $\sigma_{\lambda}$ is an unknown.
Therefore, to perform our analysis we prescribe a line width to the prominence based on the thermal velocity of hydrogen at $8000$\,K (the influence of other values is investigated in Appendix \ref{line_width}).
In future, we would suggest that a scan of the line (about six points should suffice) was performed before any prominence Dopplergram observation to allow for a calibration of the average line width that is more accurate.

\section{General properties of the prominence velocity distribution}

Before we present our investigation of the correlations in the prominence velocity field, we first provide some basic information about the characteristics of the velocity distributions of the prominence.
Figure \ref{prom_data} gives the log mean intensity, mean velocity, standard deviation of the velocity and correlation time of the velocity in the prominence.
The figure shows that values between 200\,s and 500\,s are common for the correlation time.
The longest correlation time found is 9000\,s, which is significantly shorter than the time series of the observations.
The mean velocity of all the prominence pixels is $0.9$\,km\,s$^{-1}$ and the standard deviation of the velocity fluctuations {taken over both space and time} {is $v_{\rm RMS}=2.0$\,km\,s$^{-1}$.}

\begin{figure*}[ht]
\centering
\includegraphics[width=7.5cm]{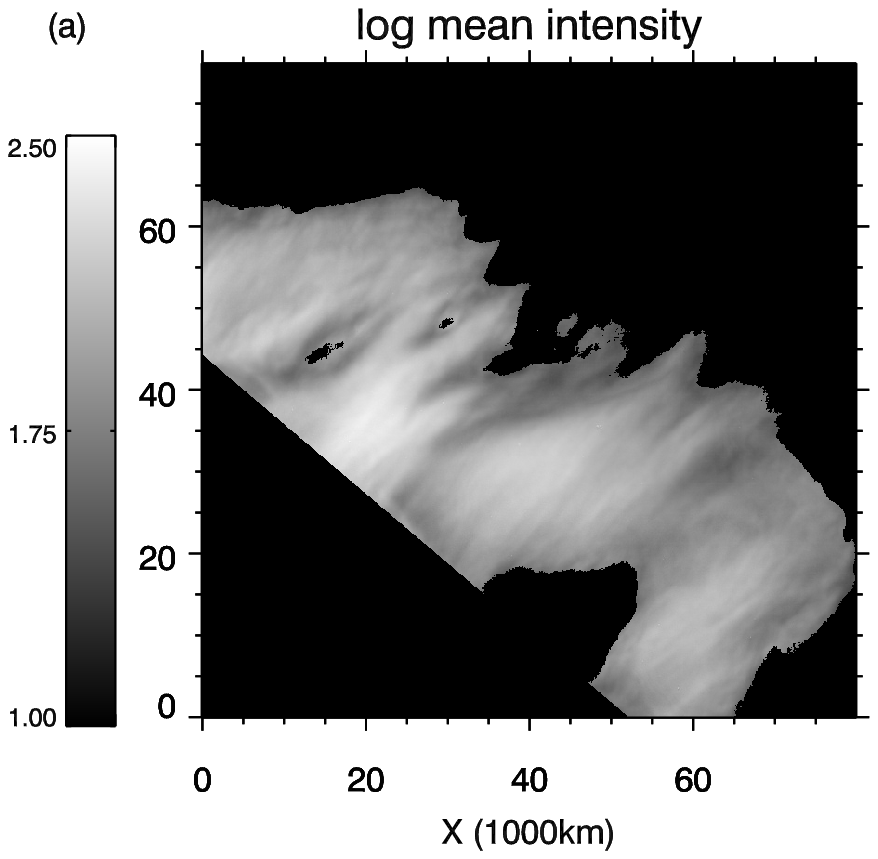}
\includegraphics[width=7.5cm]{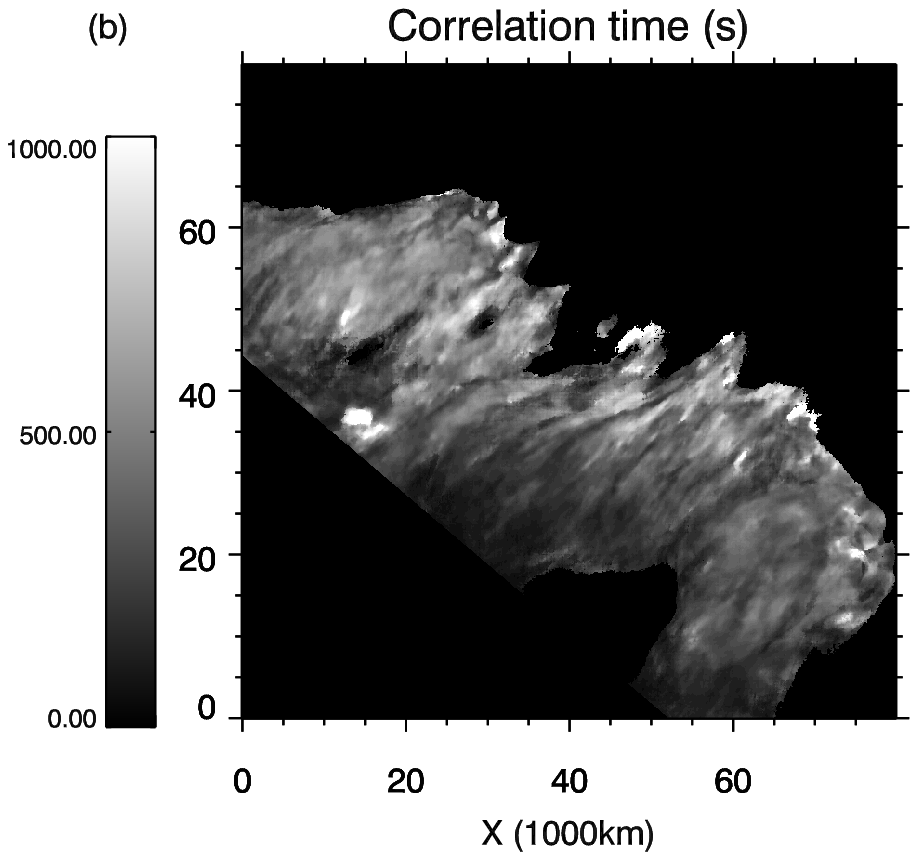}\\
\includegraphics[width=7.5cm]{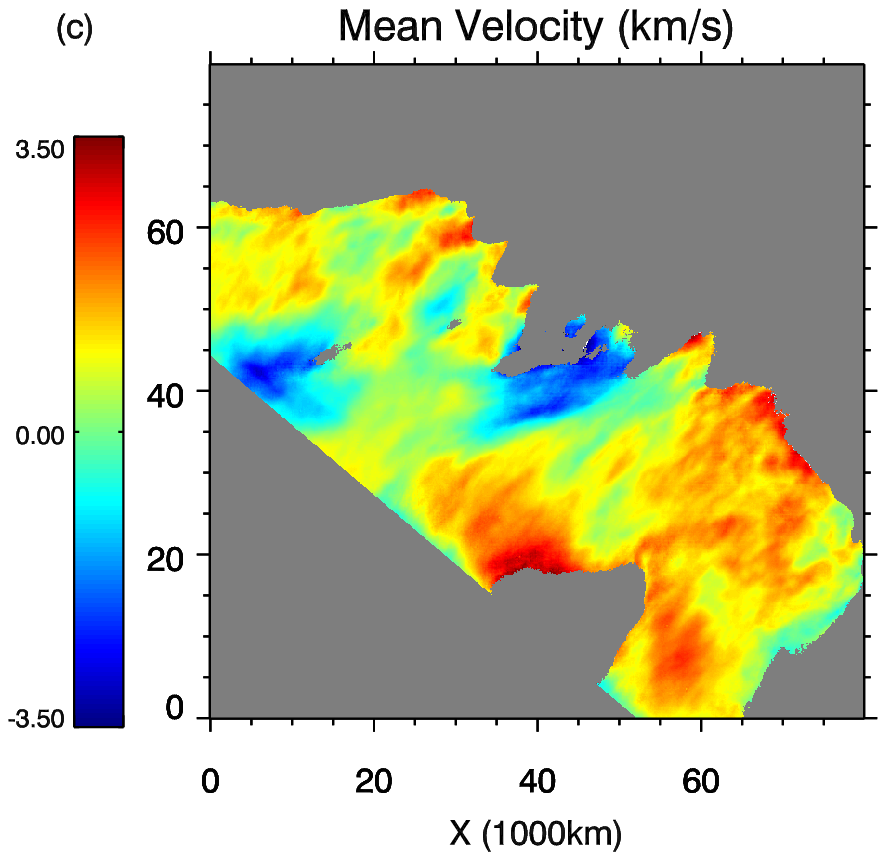}
\includegraphics[width=7.5cm]{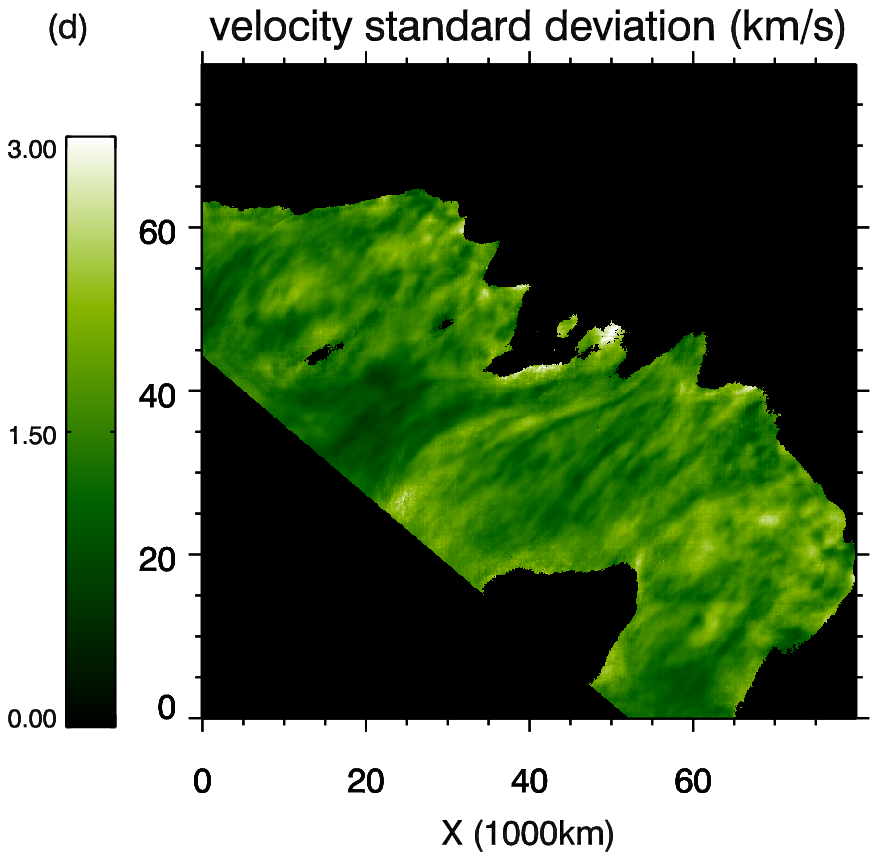}
\caption{{In panel (a) the log of the temporal mean of the sum of the intensity from both wings is given. Panel (b) shows the map of the correlation time calculated as the {Half-Width Half-Maximum (HWHM)} of the auto-correlation function of the prominence velocity fluctuations. Panels (c) and (d) respectively give the {temporal} mean velocity and standard deviation of the velocity at each pixel of the prominence.}}
\label{prom_data}
\end{figure*}

Figure \ref{acf} shows {the normalised average at each lag of the auto-correlation for all} the prominence pixels for both the velocity and the total intensity.
From this we can see that the correlation time for the velocity {(as calculated from the {half-width half-maximum (HWHM)} of the auto-correlation) is $328$\,s and for the intensity it is $544$\,s.}
We can compare this with the values found for the motions of the solar photosphere where \citet{MK2010} found that the g-band intensity correlation time was $250$\,s but the velocity correlation time was $100$\,s.

\begin{figure}[ht]
\centering
\includegraphics[width=8cm]{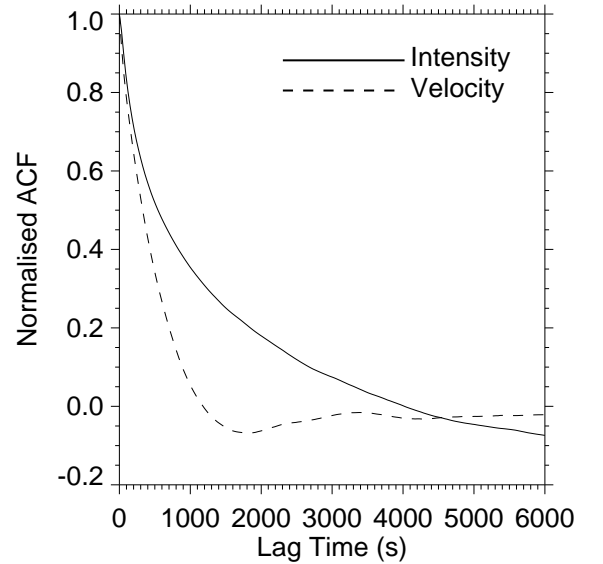}
\caption{Normalised average at each lag of the auto-correlation function for all the pixels of the prominence for both the prominence intensity (solid line) and velocity fluctuations (dashed line).}
\label{acf}
\end{figure}

Figure \ref{regions} shows the velocity distribution at 12:37\,UT for the prominence.
The velocity map of the prominence shown in this figure has been rotated so that the y-axis is aligned with the vertical direction (i.e. aligned with the local gravity) and as such the x-axis can be seen as the horizontal direction.
These two directions will form a key part of the analysis presented in this paper.
The three boxed regions have been selected because of the different characteristics that can be found.
Region 1 (R1) is a region of the prominence that is relatively quiescent.
Region 2 (R2) is a region that displays regular formation of downflows similar to those studied by \citet{Chae2010}.
Region 3 (R3) is a region that has multiple plume formation through the Rayleigh-Taylor instability as investigated by \citet{BERG2010}.

\begin{figure*}[ht]
\centering
\includegraphics[width=16.5cm]{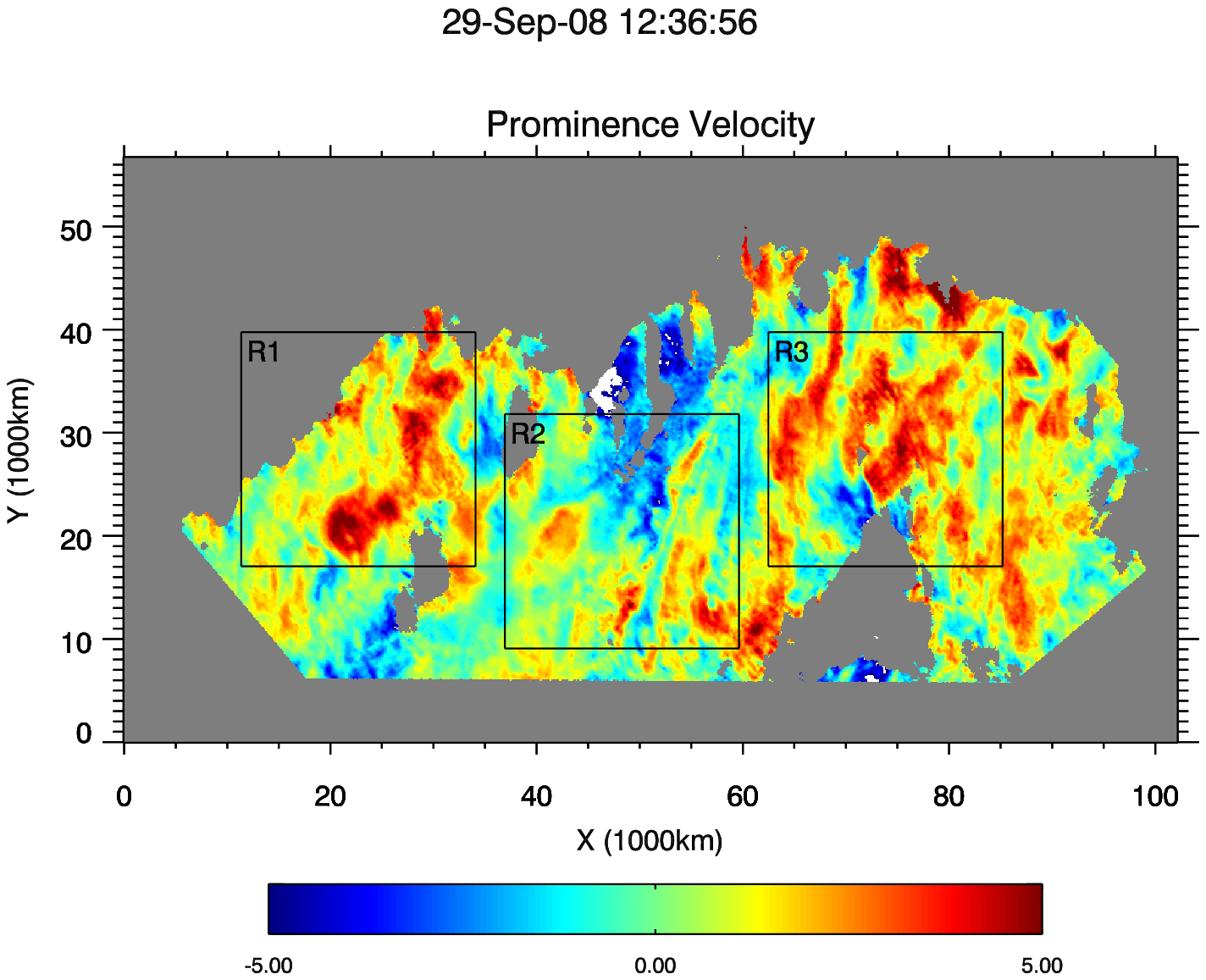}
\caption{Velocity map of the prominence at 12:36:56 UT. The three regions used in the investigation are marked by the three boxes. Online movie available for this figure.}
\label{regions}
\end{figure*}

Figure \ref{histo} shows the histograms for the velocity for R1, R2 and R3 and the whole prominence.
The solid black line gives a Gaussian pdf.
Generally speaking, when the velocity is less than two standard deviations from the mean ($|v-\mu_R| <2\sigma_R$) the distributions are close to Gaussian, but departures from a Gaussian distribution are found in the wings of the distribution.
Table \ref{table1} gives the key parameters associated with each distribution.

\begin{figure}[ht]
\centering
\includegraphics[width=8cm]{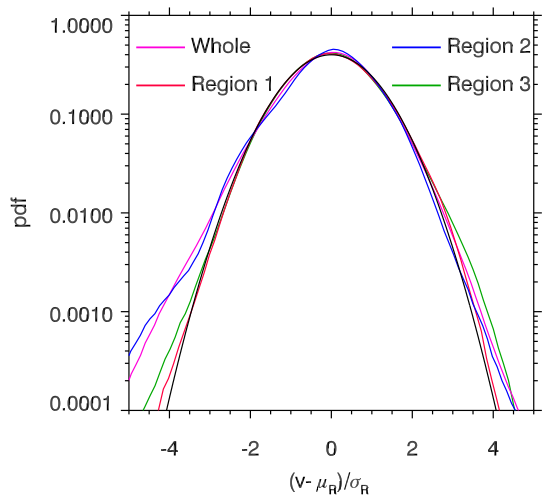}
\caption{Pdf of the velocities for the whole prominence and the three regions, where $\mu_R$ is the sample mean of the velocity for each given region and $\sigma_R$ is the standard deviation of each given region and these are used to normalise each distribution (see the second and third columns of Table \ref{table1}). The black line shows a Gaussian pdf.}
\label{histo}
\end{figure}

 
\begin{table*}
\begin{center}
\caption{Parameters of the velocity distribution of the whole prominence and the three regions.}
\begin{tabular}{ c  c  c  c  c  c }
\hline
Region & Vel. mean (km\,s$^{-1}$)  & Vel. sd (km\,s$^{-1}$) & Skew & Kurtosis & Correlation time (s) \\ \hline
Whole  & $0.9$ & $2.0$ & $-0.18$ & $0.89$ & $328$   \\
R 1    & $0.8$ & $1.7$ & $0.01$  & $0.07$ & $494$   \\
R 2    & $0.4$ & $1.7$ & $-0.42$ & $0.46$ & $246$   \\
R 3    & $1.2$ & $1.6$ & $0.05$  & $0.50$ & $317$   \\
\hline 
\end{tabular}
 
\label{table1}
\end{center}
\end{table*}

The key results of Table \ref{table1} can be summarised as follows:
\begin{enumerate}
\item R3 has the highest mean velocity of $\sim1.2$\,km\,s$^{-1}$ and R2 has the smallest of $0.4$\,km\,s$^{-1}$
\item The standard deviation of the velocity is very consistent across the whole prominence.
\item R2 has the strongest skew. This could be associated with the large number of impulsive downflows that are observed to have a unidirectional Doppler signal toward the observer. This is not so prevalent in R3 as the Rayleigh-Taylor dynamics present a broader spectrum of velocities around the plume head \citep{OROZ2014}.
\item Most regions have strong positive kurtosis $ K(v) =  \langle (v - \langle v \rangle)^4 \rangle / \langle (v - \langle v \rangle)^2 \rangle^2 - 3$. Though R1 has a relatively small kurtosis compared to the other regions, potentially {this results} from the more quiescent nature of this region.
\item The correlation time is shortest in R2. This could be a result of the downflow activity.
\end{enumerate}
{The values of Vel. mean in Table \ref{table1} are calculated with respect to the zero position of the tunable filter. 
This, however, may not correspond to shifts from the rest wavelength at the solar limb due to the difficulties of accurately calibrating the tunable filter.
Therefore, relative velocity values (as investigated from this point in this paper) can be used without any issue, but the absolute magnitude of a given velocity may not correspond to that of the shift from the rest wavelength.}

\section{Properties of the velocity correlations}

Here we present the investigation of the correlations of the velocity field, as revealed by an analysis of the structure functions.
The whole prominence and R1, R2 and R3 are investigated.

\subsection{The whole prominence}\label{whole_prom}

We first investigate the velocity correlations in the whole of the prominence to look for basic information on the nature of the turbulent flows in the prominence.
The two signs that we will look for are the progressive development of non-Gaussian tails as we investigate the pdfs of the velocity increments ($\delta_r v$) for decreasing separation ($r$) and self-similarity, that is showed by the existence of power laws in the structure functions of the velocity increments over an extended range of separations.
These statistics are performed using temporal separations of $3000$\,s between frames used for calculating the structure functions (N.B. it was not necessary to take such a large separation, in fact using any separation that is greater than the correlation time would suffice, however the large number of data points for even this separation made it sufficient).

\begin{figure*}[ht]
\centering
\includegraphics[width=8cm]{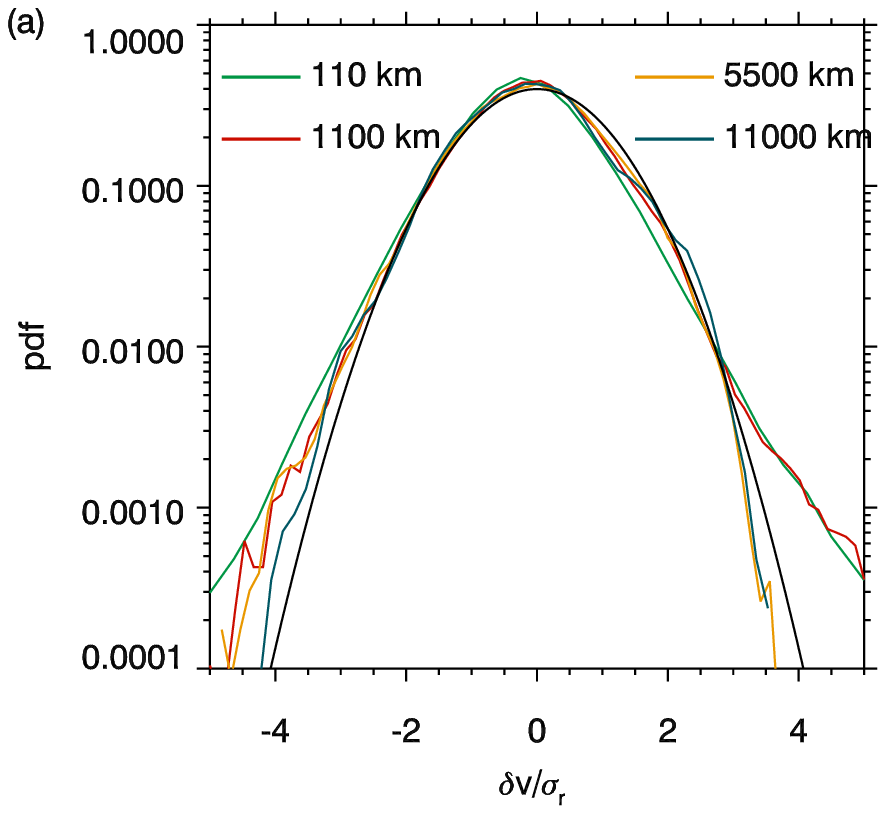}
\includegraphics[width=8cm]{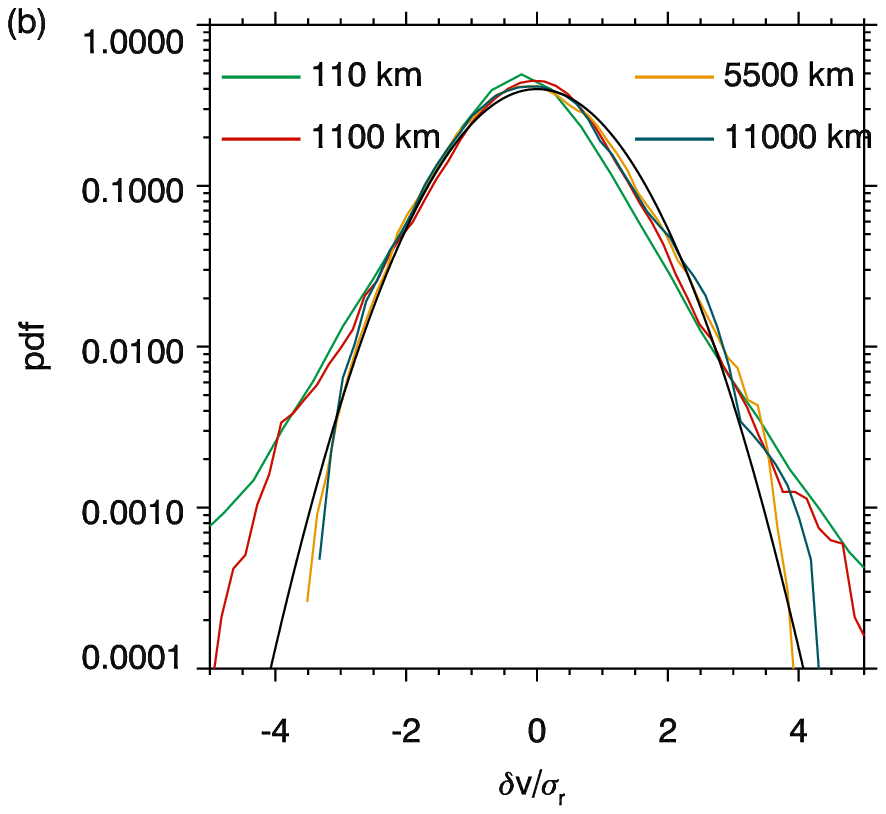}
\caption{Pdf of the velocity increments at separations of $110$\,km, $1100$\,km, $5500$\,km and $11000$\,km with the black line giving the distribution of a Gaussian pdf. Panels (a) and (b) given the distributions for horizontal and vertical separations respectively.}
\label{histo_1_order}
\end{figure*}

Figure \ref{histo_1_order} gives the probability density function (pdf) of the {velocity increments for different separations for both the horizontal direction (panel a) and the vertical direction (panel b).
The black line, given as a reference, is the pdf of a Gaussian distribution.
From this figure it is clear that at larger separations the distribution is approximately Gaussian. 
Hence fluctuations are likely to be largely uncorrelated.
However, as scales get smaller and smaller the tails of the pdfs become more and more non-Gaussian.
This behaviour has been shown to be associated with intermittency {\citep{FRIS1996}} in turbulence and provides one piece of evidence that the observed velocity field of the prominence is intermittent.

\begin{figure*}[ht]
\centering
\includegraphics[width=16.5cm]{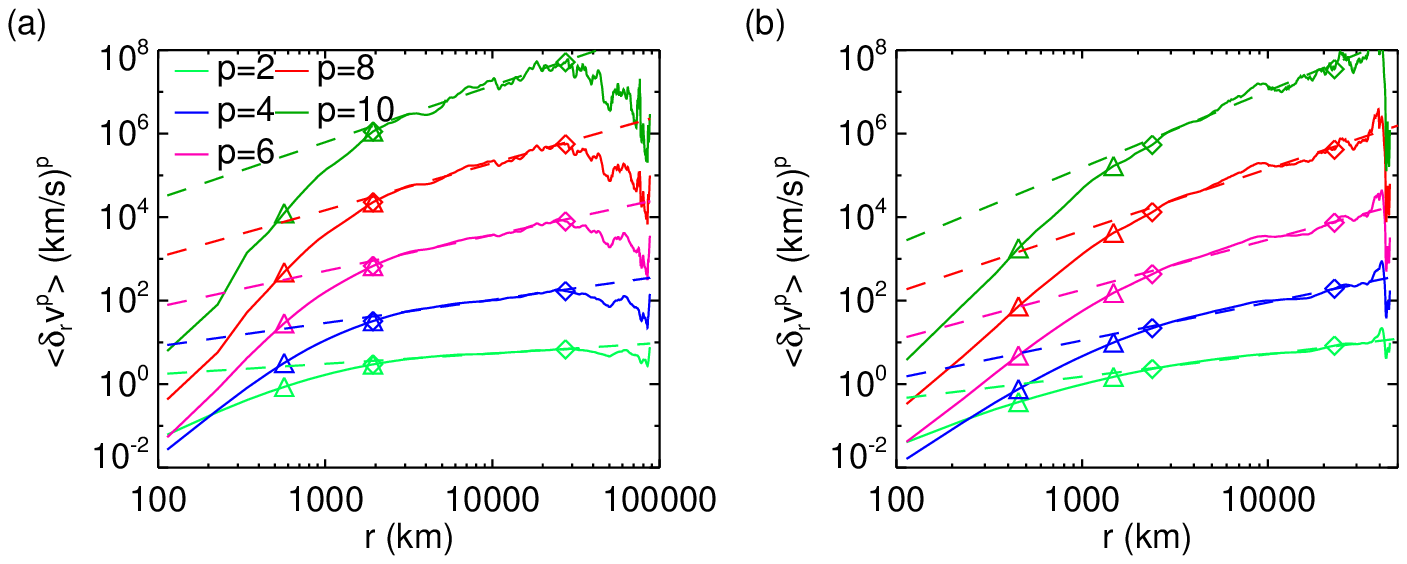}
\caption{Even order structure functions between $p=2$ to $10$ across the whole prominence for the velocity increments calculated in the horizontal (panel a) and vertical (panel b) directions. {The triangles and diamonds mark the ranges where the exponents of the power law are calculated.}}
\label{big_2_order}
\end{figure*}

{The structure functions $\langle \delta_r v^p \rangle$ against $r$ for the even values of $p$ between $p=2$ and $p=10$ are given in Fig. \ref{big_2_order} for both the horizontal (panel a) and vertical (panel b) velocity correlations.}
The first point {to note} is that the distribution seems to fall into four separate {ranges} as defined by certain lengthscales.
The first {range}, as associated with the smallest of observable {lengthscales, that is those less than $ 500$\,km, can} be associated with the finite resolution of the observations.
The second {range} spans approximately  between $500$\,km and $2000$\,km and shows a power law, this range is marked by the two triangles.
The third {range} spans approximately between $2000$\,km and $3\times10^4$\,km and shows an elongated power law region whose scaling exponent is smaller than that of smaller separations, this range is marked by the two diamonds.
{In the vertical scalings as shown in Fig. \ref{big_2_order} (b) there is some hint of a fifth short range between $\sim 1000$\,km and $\sim 3000$\,km
This {may be} a result of the scales associated with the multitude of upflows and downflows in the prominence, for example multiple downflowing knots or increasingly elongated Rayleigh-Taylor plumes, blurring the transition between the two exponents of the power law.
This transition where the exponent of the power law changes (i.e. approximately $2000$\,km) is consistent with the lengthscale where there exists a knee in the power law for the power spectral density of both the intensity fluctuations \citep[e.g. Fig 3. of][]{LEO2012} and for the plane-of-sky velocity field \citep[e.g. Fig. 4 of][]{FREED2016} of prominences.}
The fourth and final {range} starts at approximately $3\times10^4$\,km.
As we show in the next subsection, the separation scale at which the fourth region begins is dependent on the size of the region being investigated and so is likely to be a {result of smaller} sample sizes of velocity increments that are available at larger separations.

\subsection{The three regions}

Now we will look at the three separate regions of the prominence to see if the different dynamics found in the prominence influence the statistics. 
To reduce spurious correlations that may arise in the temporal averaging process, we include only the data for the calculation of the velocity increments using snapshots taken at the time separation nearest to the correlation time for each region as listed in Table \ref{table1}.

\begin{figure*}[ht]
\centering
\includegraphics[width=16.5cm]{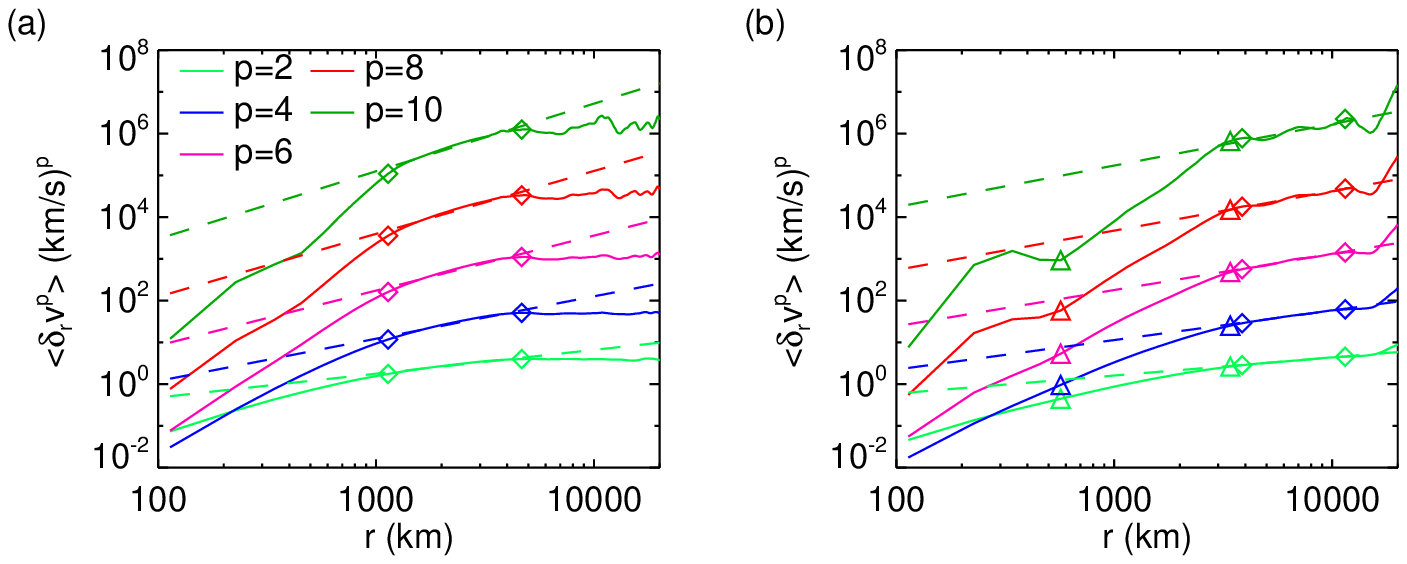}\\
\includegraphics[width=16.5cm]{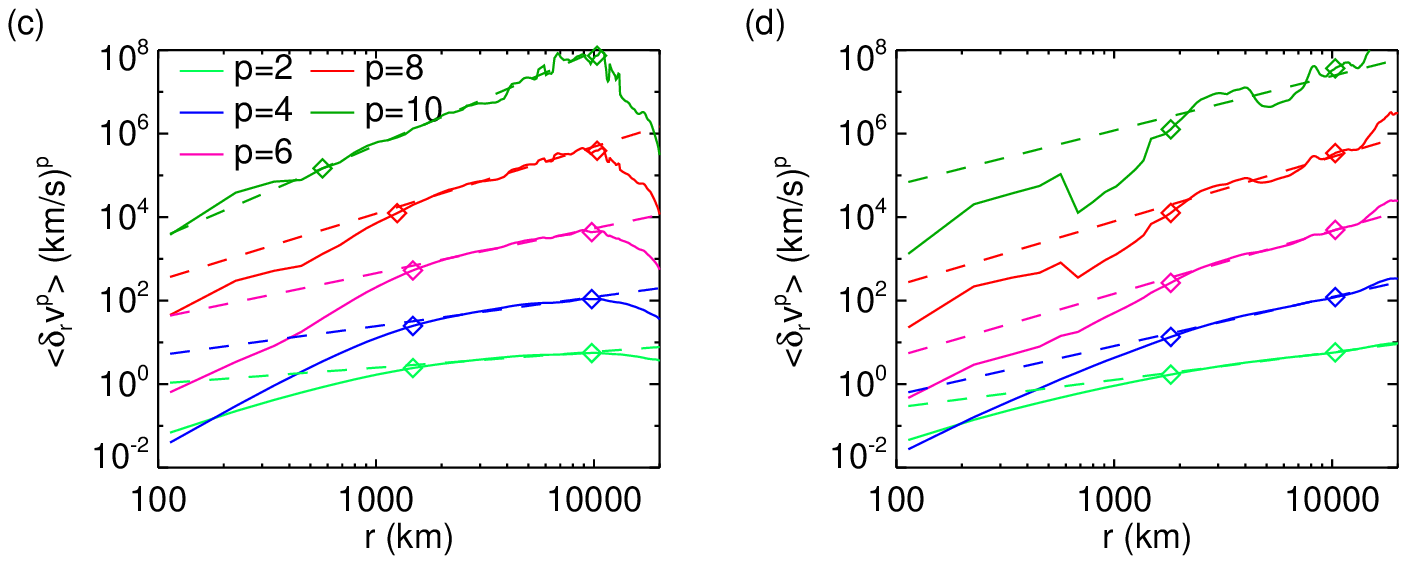}\\
\includegraphics[width=16.5cm]{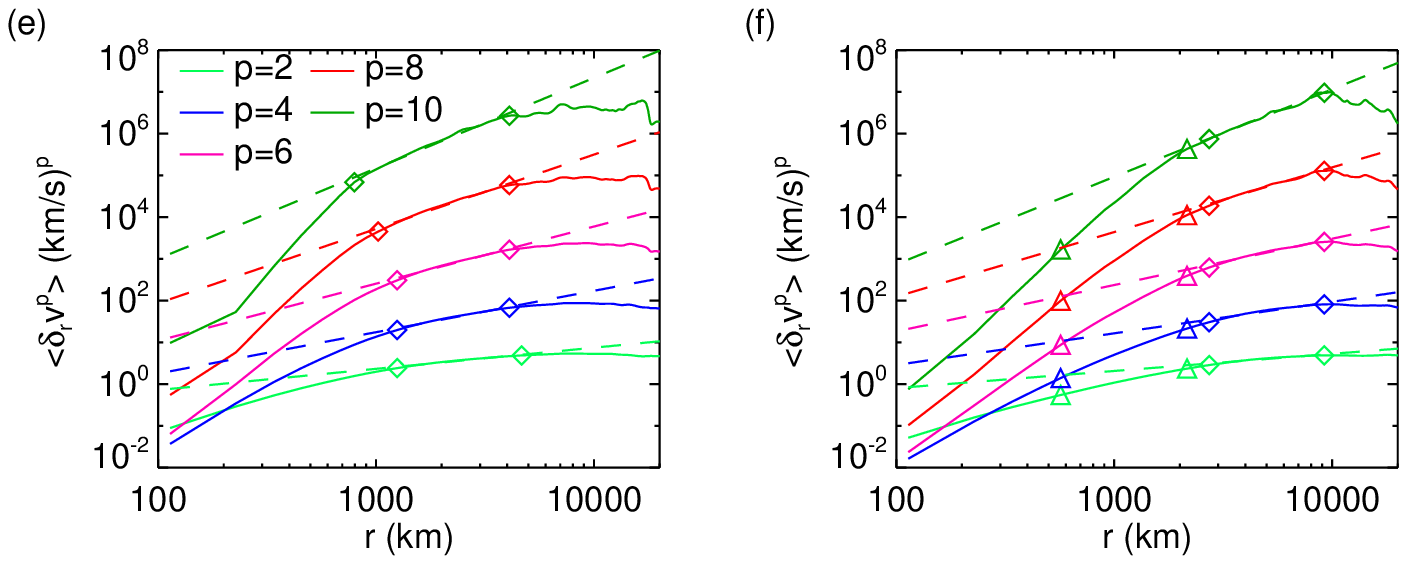}
\caption{Even order structure functions between $p=2$ and $p=10$ for the three regions of the prominence as shown from top to bottom: in panels (a) and (b) R1, in panels (c) and (d) R2 and in panels (e) and (f) R3 from Fig. \ref{regions}. {The left and right panels give the structure functions calculated from the horizontal and vertical separations respectively. The triangles and diamonds mark the ranges over which the exponent of the power law are calculated.}}
\label{3regions_highorder}
\end{figure*}

Figure \ref{3regions_highorder} shows the even order structure functions of $\langle \delta_r v^p \rangle$ between $p=2$ and $p=10$ for the three regions of the prominence as shown from top to bottom.
The general trends found are similar to those shown in Fig. \ref{big_2_order}.
One interesting point that should be noted is the difference between the lengthscale associated with horizontal and vertical break in the power law, though the position of the break in the horizontal power law is relatively similar for the three regions, this is not the case for the vertical scaling.
The approximate position of the break in the vertical power law for R1 happens at $4000$\,km, for R2 at $2000$\,km and for R3 at $2500$\,km.
This could be related to the different dynamics of the regions, where R2 {and} R3 are dominated by vertical flows but R1 may be dominated by the vertical threads that are often observed in prominences.
Because the lengthscale at which the break in the power law occurs for R1 has been pushed to longer lengthscales, this allows for the range of the steeper power law to be observed over an extended range providing greater evidence of its existence.

Figure \ref{time_struct_fun} gives the {second-order} temporal structure function, $\langle \delta_t v^2 \rangle = \langle [v(\mathbf{x}, s + t) - v(\mathbf{x}, s)]^2 \rangle$ against $t$ for the three regions.
For the temporal distribution, power laws exist only for temporal separations below $\sim 1000$\,s.
The exponents for the three regions are approximately $1$ for R1, $0.6$ for R2 and $0.9$ for R3.
{It is interesting that the exponent is noticeably reduced for R2 compared to the other regions.
This implies that this region has a higher ratio of kinetic energy at the higher frequencies to the lower frequencies than the other regions. 
As can be seen in the movie associated with Fig. \ref{regions}, this region is dominated by the downflowing prominence knots.
Also, it is interesting that the shorter the correlation time for a region, as given in Table \ref{table1}, the smaller the exponent.}
Above the temporal separation of approximately $1000$\,s the distribution is almost flat for all three regions.
{One potential physical meaning of this value is of the period of the Alfv\'{e}n waves or the timescales of the dynamics that drive the turbulence.}

\begin{figure}[ht]
\centering
\includegraphics[width=8cm]{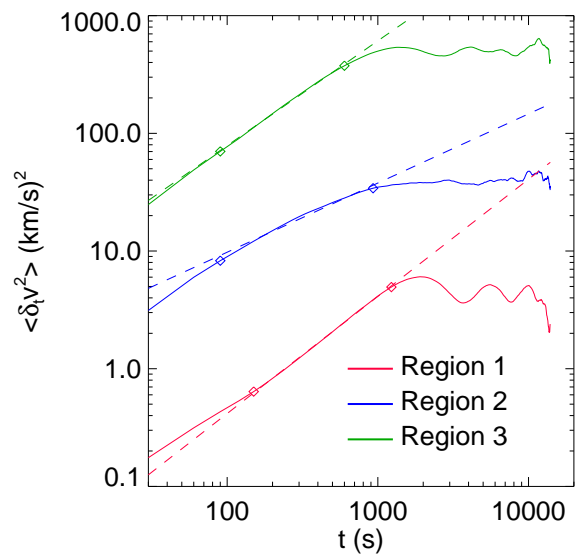}
\caption{Plot of $\langle \delta_t v^2\rangle$ against $t$ for R1, R2 \& R3. The exponents of the power laws for the three regions are approximately $1$ for R1, $0.6$ for R2 and $0.9$ for R3. Note that the distributions are scaled to make the distribution clear.}
\label{time_struct_fun}
\end{figure}

{It is interesting to try to connect the temporal to the spatial structure functions.
One way to do this would be to use Taylor's hypothesis that relates the temporal to spatial scales with $r=U t$ where $U$ is the magnitude of the mean flow velocity.
As can be seen in the movie of Fig. \ref{regions}, there is no clear particular mean flow in the plane-of-sky that can be determined and so the assumption of the Taylor hypothesis that $U$ when compared to the turbulent velocity $u_{\rm turb}$ satisfies $u_{\rm turb}/U\ll 1$ is not valid, but even so it is worth using this simple scaling to see if it provides any information.
Here we have $\langle \delta_t v^2 \rangle \propto t^{0.5}$ to $t^1$ which would map to $\langle \delta_r v^2 \rangle \propto r^{0.5}$ to $r^1$ which are exponents that are consistent with those found for the spatial separations (see Fig \ref{power_vs_order}).
However, this does not allow us to determine if the power law found in the second-order temporal structure functions is related to the distance structure functions.
However, comparing the time that relates to the knee in the temporal structure functions ($t\sim1000$\,s) to the lengthscale that relates to the knee in the spatial structure functions $r\sim2000$\,km gives a speed $S=2000/1000=2$\,km\,s$^{-1}$ which is comparable to $v_{\rm RMS}$ of the prominence.}

\subsection{Looking at the higher order structure functions}

In Figs. \ref{big_2_order} and \ref{3regions_highorder} we show the even order structure functions against separation $r$, but it is important to understand what being shown in these higher orders.
Figure \ref{high_order_pdf} shows the pdf of the velocity increments at a separation lengthscale of $3000$\,km multiplied by ($\delta_r v/5\sigma_r)^n$ for even numbers between $n=2$ to $10$.
{We note that the factor five is only used to rescale the distributions and has no physical meaning.}
The integral across this whole distribution then gives the value of the structure function for this separation, so we are able to understand which values of velocity separation are contributing most to which order of structure function.
It can be seen that as we go to higher orders of the structure functions the structure function samples further into the wings of the $\delta_r v/\sigma_r$ distribution.
From the left panel of Fig. \ref{high_order_pdf}, we empirically conclude that the moments of the horizontal velocity increment is reliable upto the order 6th (the 10th moment is not and the 8th is marginally reliable). 
From the right panel of Fig. \ref{high_order_pdf}, we conclude that the moments of the vertical velocity increment is reliable up to the order 4th (the 8th and 10th moments are not and the 6th is marginally reliable).

\begin{figure*}[ht]
\centering
\includegraphics[width=18cm]{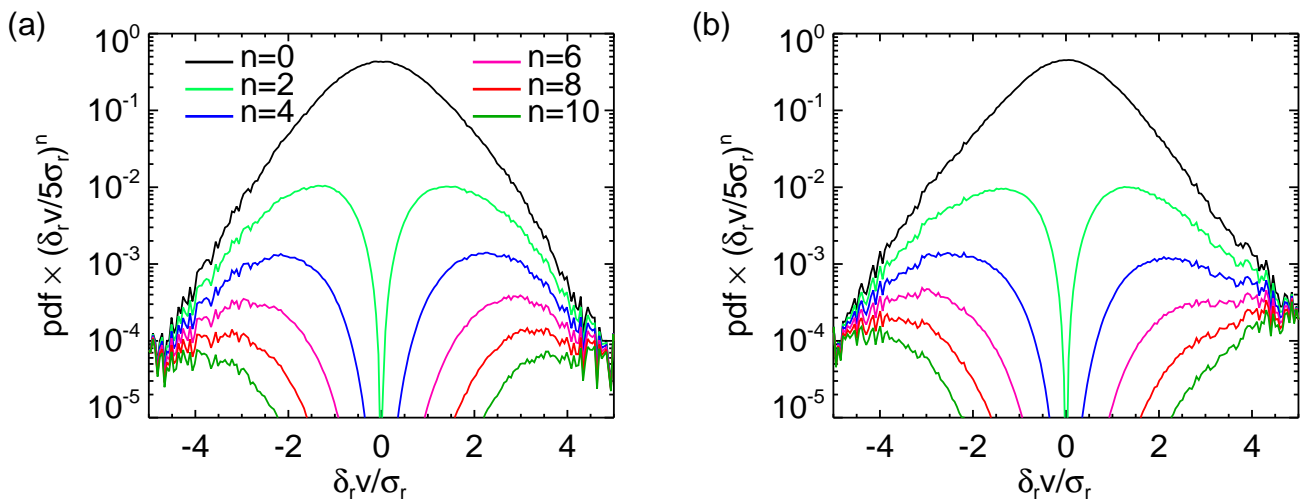}\\
\caption{{Pdf of} the velocity increments at a separation lengthscale of $r=3000$\,km multiplied by ($\delta_r v/5\sigma_r)^n$ to show which parts of the pdf distribution are sampled by the different order structure functions. The case where $n=0$ gives the reference pdf. Panel (a) relates to the velocity increments calculated from horizontal separations with panel (b) being the same but calculated from vertical separations.}
\label{high_order_pdf}
\end{figure*}

In Fig. \ref{3regions_highorder} we can {see for the horizontal scalings of R2 and R3 that the position of the lengthscale for the break in the power law shifts to smaller lengthscales with increasing $p$ (see panels c and e of Fig. \ref{3regions_highorder}).
{Such} a break can be hypothesised to exist at a lengthscale when two physical processes with timescales $\tau_1(r)$ and $\tau_2(r)$ exists such that $\tau_1(r)=\tau_2(r)$, for example for weak and strong MHD turbulence when $\chi_r=1$ as defined in Equation \ref{nonlin}.
If the intermittency is only weak, and the timescales at a given lengthscale {are given by $\tau^p=r^p/\langle \delta v_r^p \rangle$} then the lengthscale where the break appears should remain the same at all values of $p$.
However, if the intermittency is high then relation between each timescale and the lengthscale changes resulting in the lengthscale at which the transition occurs shifting.
Therefore, this could be a very interesting physical phenomenon to investigate further, but as we can see in Fig. \ref{high_order_pdf} the highest order structure functions have a significant contribution coming from regions with high proportion of noise.
To investigate whether this is a real phenomenon, more events would be required at these extreme values.}

Figure \ref{power_vs_order} gives the result from calculating the exponents of the two power law distributions found in the prominence structure functions for {different orders} $p$ of the structure function.
{The exponents calculated from the structure functions for the horizontal (solid line) and vertical (dashed line) velocity increments for the three regions and the whole prominence from both above and below the break in the power law are included.}
Two clusterings of the values of the exponents can be seen when looking at order $p=2$.
This is $1=p/2$ and $0.5=p/4$.
The clustering around $1$ is associated with the power law at lengthscales less than $2000$\,km and the clustering around $0.5$ is associated with the power law at lengthscales greater than $2000$\,km.
{Both of these clustering indicate that the exponent is a nonlinear function of $p$. This is another signature of intermittency of the velocity fluctuations. The lower and upper solid black lines show what would be expected if the exponents follow the Kraichnan-Iroshinikov scaling, Eq.(\ref{KIspec}), and the weak-MHD-turbulence scaling, Eq.(\ref{weak_MHD}), respectively.}

{There are two possible errors we consider as the source of errors shown in Fig. \ref{power_vs_order}.
First the error ($\sigma_{fit}$) found in the slope (i.e. the exponent of the power law) given by a linear fit to the log of $\langle\delta_rv^p\rangle$ over the specified range. 
However, this was found to not reflect the lower reliability of the exponents of the higher order structure functions.
To reflect this, the exponent at each snapshot of the data used was then calculated and from this it was possible to calculate the standard error (the standard deviation divided by the root of the number of values) of the exponent for each snapshot ($e_S$) with respect to the exponent calculated for all data ($e_{\mu}$) giving $\sigma_{\rm SE}=\sqrt{\Sigma(e_S-e_{\mu})^2}/S$, where $S$ is the number of snapshots.
The error in Fig. \ref{power_vs_order} is given as $\sigma=(\sigma_{fit}^2+\sigma_{\rm SE}^2)^{1/2}$.}

\begin{figure}[ht]
\centering
\includegraphics[width=9cm]{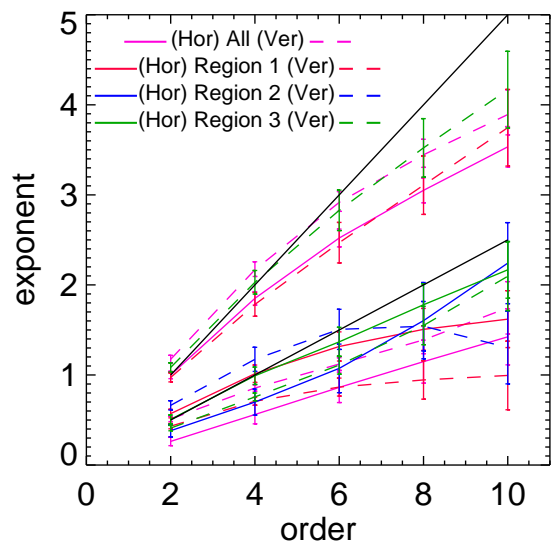}
\caption{Scaling exponent of the structure function as a function of the order $p$ both above and below the break in the distribution for the whole prominence and all three regions, both {for} horizontal (solid lines) and vertical (dashed lines). The black lines give the expected trend if the exponents of structure functions followed the linear relation of $p/2$ (higher) or $p/4$ (lower). {Note that below the break, only some of the structure functions displayed a clear power law for all orders of the structure function and so only four exponents are plotted from below the break.}}
\label{power_vs_order}
\end{figure}


\section{Discussion}

In this paper, we have presented the analysis of turbulent fluctuations of the line-of-sight velocity field of a quiescent prominence.
Using structure functions to analyse the velocity field, we find that as we look at smaller spatial separations there is an increasing departure from Gaussianity for the pdf of the velocity structure functions and that there are two separate power laws with exponents roughly consistent with $p/2$ at small scales and $p/4$ at large scales for the spatial distribution with the break found at approximately $2000$\,km and one power law found in the temporal distribution for temporal separations less than $1000$\,s.
We have also presented new techniques for the data reduction and analysis of Hinode SOT H$\alpha$ Dopplergrams.

\subsection{{Explaining the power law exponents and the break in the power law}}

{One of the key area that should be focussed on is the exponents of the power law and the existence of the break in the power law at approximately $2000$\,km found in the spatial structure functions.}
Above this point, we find a scaling consistent with K-I turbulence (i.e. $\sim p/4$) or with the scaling perpendicular to the magnetic field in some models of strong MHD turbulence \citep{BOLD2005}, but below this point the scaling is consistent with weak MHD turbulence (i.e. $\sim p/2$).
As we have explained in the introduction {with the nonlinearity parameter (see Equation \ref{nonlin})}, in current MHD turbulence theory it is expected that the transition between weak and strong turbulence should exist, but it is expected that weak turbulence exists at the larger scales and strong turbulence at the smaller, that is the strong turbulence scaling should exist at smaller lengthscales than the weak MHD turbulence scaling, which is the opposite of what we have found.
Therefore, it is likely that something about the prominence system under study, or more likely the prominence system in general, is creating this complexity.

The first explanation that should be investigated is {the change in exponent of the power law} being an artefact of the observations {or the analysis techniques used.}
Considering that the break in the power law appears at scales over an order of magnitude larger than the pixel size, {we can discount insufficient resolution as a reason and} a more complex explanation than the scales being unresolved has to be invoked.
The data that we use to create the Dopplergrams, and ultimately the velocity map, comes from two images {taken in} the blue and red wing of the H$\alpha$ line with a 10\,s gap between the two images. 
If some information could be transferred between different positions in the prominence during this time, by waves for example, then the break in the power law could {occur as a result}.
However, this would imply a wave velocity of $v_{\rm wave}\sim 2000$\,km$/10$\,s$=200$\,km\,s$^{-1}$, which is much faster than the expected wave speeds for {a quiescent prominence.
Also, a similar behaviour of a power law with a break at scales $\sim 2000$\,km was also found on different prominences using very different analysis techniques, by \citet{LEO2012} and \citet{FREED2016}, where these were found in the Ca II H data, which does not have this issue with a 10s delay and also shows how the choice of spectral line is not critical for finding this behaviour.
This gives us confidence to say that the change in exponent is not a result of the use of structure function analysis.
Based on this evidence, we conclude that there is no obvious reason that the break in the power law is a result of some observational or analysis artefact.}

One explanation for the {change in the exponent in the power law} could be that at smaller scales the turbulence may be {as a result of} local excitation in the prominence, but at larger scales the relation of the motions to the surrounding corona should be considered.
When thinking about this in terms of the standard prominence model, that is prominence material collects in dips of the coronal magnetic field and is supported by magnetic tension, for fluctuations on scales smaller than the scales associated with the collection of dense material the turbulence could be completely contained with the prominence material, but for scales larger than this, it could be expected that the motions are part of the global coronal-prominence system.
Now let us observe in Fig. \ref{regions} that the break scale, 2000 km, corresponds to the typical horizontal width of streaky structures such as the red finger-like high speed regions seen in the bottom of R2. 
In R1 and R2, apart from the biggest percolating structures, the typical horizontal width is also around 2000 km.
The temporal variation of the streaky structures shows vigorous fluctuations of the edges while the streaks themselves are long-lived structures. 
They are actually jets taken in the fluid dynamics sense (to be precise they are the dynamic Rayleigh-Taylor plumes, prominence knots and other impulsive flows of the system).
Thus the nature of fluctuations below 2000 km, which are dominated by a single streak, can be different from the one above which is determined collectively by multiple streaks and other large scale dynamics. 
Further, the fluctuations of the streak edges can be viewed as random waves, which is consistent to our picture of the coexistence of the weak MHD wave turbulence in the small scale and the strong turbulence in the large scale.

There is evidence for small-scale wave turbulence excited in a larger turbulent system by the Rayleigh-Taylor instability in a different area of fluid dynamics.
\citet{CHER2005} presented a phenomenological model where a wave energy cascade (in this case surface capillary waves) propagating along the surface of the rising bubble could be formed.
There are some similarities between the situation they suggest and the one found here, and so this could explain the puzzling break in the power law of the prominence.
{However} we notice that in our hypothetical wave turbulence scales the pdf of the velocity increment is not Gaussian as indicated in Fig. \ref{histo_1_order} (1100 km case), which is contrary to the near Gaussian behaviour expected of wave turbulence in general.

Some evidence for the connection between the power laws found at large scales in the prominence and the coronal turbulent motions may be present in the CoMP observations of the Doppler velocity spectra for trans-equatorial coronal loops by \citet{TOM2009}.
The power law for the power spectral density of the frequency was found to be $\sim \nu^{-1.5}$, which is equivalent to $\langle\delta_t v^2\rangle\sim t^{0.5}$ for the second order structure function of the temporal fluctuations. 
We find $\langle\delta_r v^2\rangle\sim r^{0.5}$ in {the spatial structure functions for $r$ greater than $2000$\,km, which may} hint at some connection between these two systems.
However, different from the observed coronal spectrum, the prominence temporal spectrum does not show any evidence of the p-mode excitation and the exponents observed are different, other than for region 2. 


\subsection{Estimation of turbulent heating}\label{heat_sec}
One very important task when assessing the turbulence in a system is to measure the energy dissipation rate of the system.
For prominences, this will tell us whether energy dissipation as part of the turbulent cascade is an important part of the heating and cooling processes that occur in prominence plasma.

When we are not looking at the dissipation range of turbulence, it is still possible to calculate the amount of energy dissipated by calculating the amount of energy that is reaching the dissipation scale.
This comes from the calculating the energy transmission rate $\epsilon$ of the turbulent cascade and for this purpose the third order structure function of the velocities aligned with the separation vector $r$, which investigates the transfer of kinetic energy between different spatial scales, is of great importance.
For hydrodynamic turbulence one of the key results is known as Kolomogorov's $4/5$ law, given by \citep[e.g.][]{Davidson}:
\begin{equation}\label{45law}
\langle[(\mathbf{v}(\mathbf{x}+\mathbf{r})-\mathbf{v}(\mathbf{x}))\cdot \hat{\mathbf{r}}]^3  \rangle=-\frac{4}{5}\epsilon r,
\end{equation}
where $\hat{\mathbf{r}}$ is the unit vector in the direction of the separation $\mathbf{r}$.
Here the sign of the term on the right-hand side is showing that the energy is cascading from large to small scales.
To apply this to the prominence system, we need to think about how this would work in a system where the role of the magnetic field is crucial.

For the prominence under study, we have found two regions with two different power laws, with the power law at the smaller separations consistent with weak MHD turbulence.
For weak MHD turbulence, the third order structure function of the form presented for the $4/5$ law is given as:
\begin{equation}\label{mhd45law}
\langle[(\mathbf{v}(\mathbf{x}+\mathbf{r})-\mathbf{v}(\mathbf{x}))\cdot \hat{\mathbf{r}}]^3  \rangle \sim C''_3 \left(\epsilon \frac{V_{\rm A}}{r_{\parallel}}\right)^{3/4} r_{\perp}^{3/2},
\end{equation}
{Note that this equation has been formulated from Equation \ref{weak_MHD} taking $p=3$ and that the longitudinal velocity component (the left hand side of Equation \ref{45law}) is regarded as approximately equal to the velocity component {perpendicular} to the large-scale magnetic field, $\delta_{r\perp} v$.}
Based on this equation, we should be able to calculate $\epsilon$ for the prominence.
However, there is one problem, we do not have data to determine the full 3D velocity field as this equation requires.
To circumnavigate this issue, we will attempt to estimate the order of the energy dissipation using the following relation:
\begin{align}
\epsilon \sim & \left(\frac{\langle[v(\mathbf{x}+\mathbf{r})-v(\mathbf{x})]^3  \rangle}{r^{3/2}}\right)^{4/3}\frac{r_{\parallel}}{V_{\rm A}} = \left(\frac{\langle \delta_r v^3\rangle}{r^{3/2}}\right)^{4/3}\frac{r_{\parallel}}{V_{\rm A}}, \label{new45law}
\end{align}
where the velocities used are the observed prominence {Doppler }velocities {and the constant $C_3''$ in Equation \ref{mhd45law} is taken to be of order unity.}

\begin{figure}[ht]
\centering
\includegraphics[width=8cm]{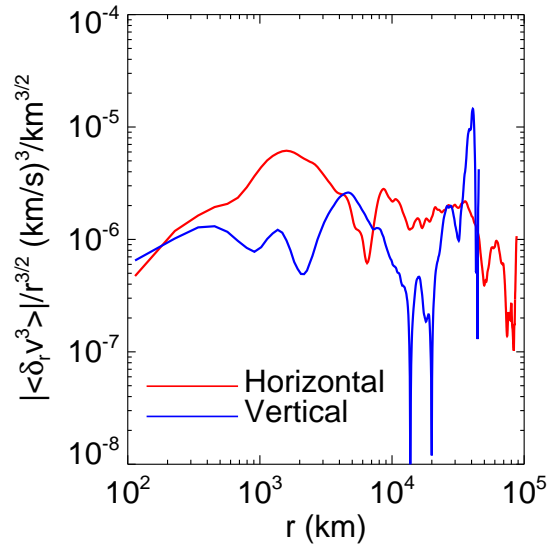}
\caption{${|\langle\delta_r v^3\rangle|}/{r^{3/2}}$  for both the horizontal (blue line) and vertical (red line) velocity increments. The modulus is taken because there are a number of $0$ crossings in the 3-rd order structure functions.}
\label{energy transfer}
\end{figure}

Figure \ref{energy transfer} shows the value of ${|\langle\delta_r v^3\rangle|}/{r^{3/2}}$ across the height and breadth of the prominence.
This value can be taken as $\sim 10^{-6}$\,$($km/s$)^3$\,km$^{-3/2}$.
{Note that at the largest values of $r$ {there} is likely to be insufficient statistics to accurately determine the values of ${|\langle\delta_r v^3\rangle|}/{r^{3/2}}$, this is equivalent to the fourth range described for the structure function scalings in Section \ref{whole_prom}, but for a range of more than two orders of magnitude the value we find holds.}
{We would like to note here that the absolute value operator is only used to keep the value plotted positive as ${\langle\delta_r v^3\rangle}/{r^{3/2}}$ can be either positive or negative.}
Taking the prominence Alfv\'{e}n speed as $20$\,km\,s$^{-1}$ {and $r_{\parallel}=20000$\,km (i.e. $20$\,km\,s$^{-1}$ multiplied by the timescale $1000$\,s - see Fig. \ref{time_struct_fun})} gives a value for the energy transmission rate of $\epsilon \sim 10^5$\,erg\,s$^{-1}$\,g$^{-1}$.
For a prominence density of $10^{-13}$\,g\,cm$^{-3}$ this would give a heating per unit volume of $10^{-8}$\,erg\,s$^{-1}$\,cm$^{-3}$.
To provide some context for this value, we can use it to estimate the time required to heat a unit volume of prominence plasma by $100$\,K.
The change in thermal energy density is given by $E_{thermal}=nk\Delta T/(\gamma-1)\sim 10^{-3}$\,erg\,cm$^{-3}$.
Therefore, it would take of the order of $10^5$\,s to raise the temperature of the prominence material by $100$\,K, which can be viewed as very inefficient heating.

\subsection{Estimation of reconnection diffusion}

{It has been established that the existence of turbulence in a magnetised medium {results} in the formation of current sheets that can lead to
reconnection in the magnetic field (see \citealt{BIS2003} and recent articles \citealt{LAZ2012a,LAZ2012b}).}
One key process that this reconnection will induce is the diffusion of mass across the magnetic field, where this process is called reconnection diffusion.
According to \citet{LAZ2012a}, the reconnection diffusion for a weakly turbulent MHD medium {is}:
\begin{equation}
\eta_{\rm rec}=v_{\rm turb}L_{\rm turb}M_{\rm A}^3,
\end{equation}
where $v_{turb}$ is the characteristic {(i.e. the value at the injection scale)} velocity of the turbulence, $L_{turb}$ is the characteristic lengthscale of the turbulence and $M_{\rm A}$ is the Alfv\'{e}nic Mach number {defined as $M_{\rm A}=v_{\rm turb}/V_{\rm A}$}.
Using the values from this study, $v_{\rm turb}=v_{\rm RMS}=2$\,km\,s$^{-1}$, $L_{\rm turb}=2000$\,km {(i.e. the lengthscale of the break in the power law) }and taking $V_{\rm A}=20$\,km\,s$^{-1}$, we can calculate $\eta_{\rm rec}\sim 4\times 10^{10}$\,cm$^2$\,s$^{-1}$.

We can also make estimates for the value of other diffusions that are present in prominences, in this case we will look at Coulomb diffusion $\eta$ and ambipolar diffusion $\eta_{\rm AMB}$.
Using a temperature of $T=10^4$\,K, an ionisation fraction of $\xi_{\rm i}=0.1$ and a magnetic field strength of $3$\,G we calculate $\eta\sim10^7$\,cm$^2$\,s$^{-1}$ and $\eta_{\rm AMB}\sim10^{10}$ -- $10^{11}$\,cm$^2$\,s$^{-1}$.
From this we can understand that the reconnection {diffusion is of} approximately the same order as the ambipolar diffusion (for these parameters at least), both of which dominate the Coulomb diffusion.
Therefore, we expect that there exists prominences such that the diffusion of neutrals across the magnetic field in the prominence should be at approximately the same rate as the diffusion of ions.

The major implication of reconnection diffusion is the transport of mass in the prominence.
As the reconnection breaks the frozen-in condition of ideal MHD, it becomes possible for mass to move through the prominence.
It was shown by \citet{PELO2005} that reconnection between two {Kippenhahn-Schl\"{u}ter} prominence model dips \citep{KS1957} results in a net flow of mass downward and a net transport of magnetic field upward.
Simulations by \citet{HILL2012b} showed that if the reconnection happens in favourable conditions, the downflowing mass would shock and that would produce the downward-propagating knots observed in quiescent prominences.
Though observations \citep{Chae2010} {suggest} that there are a large number of impulsive flows in the prominence, the analysis presented in this paper suggests that over a long time, this merely represents a slow diffusion of the total prominence across the magnetic field.

\subsection{Higher order structure functions}

Our investigation of the higher order structure functions has proved interesting where, in spite of the increase in any fluctuations resulting in a general degradation in the power laws, the same process enhanced the break in the power laws around $2000$\,km in some cases {(though it disappears in others).
We have been able to show that the} scaling exponent of the higher order structure functions, which reflect fluctuations further in the wings of the $\delta_r v/ \sigma_r$ distribution, is a nonlinear function of the order $p$ for all investigated regions of the distribution.
This is suggestive of intermittency in the inertial range of the turbulence.

In Fig. \ref{power_vs_order} we see that there is a general tendency for the exponent of the structure function to increase at slightly below that expected from a linear relation in a fashion that is consistent for both the exponents found above and those found below the break.
It is important to point out that this is consistent with the {Extended Self-Similarity (ESS) analysis presented in \cite{LEO2012} where the exponent of power law corresponding to the 2nd-order structure function as a function of the 3rd-order structure function calculated from the intensity fluctuations of a prominence gives a value of $\sim 0.7$ (where $0.66$ would be expected for a linear increase in the exponent with order of the structure function).
When we perform the same analysis with this prominences line of sight velocities, see Fig. \ref{ESS}, we find that the exponent is about $0.7$, which is generally consistent with the results of \citet{LEO2012}.}

\begin{figure}[ht]
\centering
\includegraphics[width=8cm]{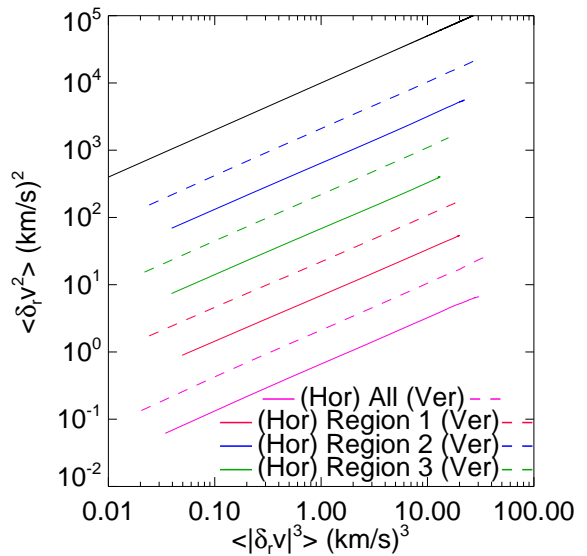}
\caption{{ESS analysis, in this case the third order structure function as calculated from the modulus of the velocity increments against the second order structure function, for the whole prominence and the three regions including both the horizontal and vertical scalings. The plots have been shifted for clarity. Black solid line gives the power law exponent of 0.7 that was found as the exponent of these plots.}}
\label{ESS}
\end{figure}

\subsection{Conclusion}

{In this paper, we have used structure functions to aid with the analysis of the line-of-sight velocity field of a prominence that has been reconstructed from Hinode SOT Dopplergrams.
{Looking at the even-order structure functions with separation $r$, we found that they display power-law scaling that are expected of turbulent media.}
However, the structure functions show that as $p$ increases the exponent increasingly deviates from the linear scaling that comes from simple dimensional analysis, implying that the system displays intermittency.
This conclusion is supported by the increasing non-Gaussianity found in the pdfs of the velocity increments when going to increasingly smaller scales.
The structure function analysis of this prominence found a break in the power law at the same scales as \citet{LEO2012} and \citet{FREED2016} found for different quiescent prominences using different methods, which may imply that this is a universal feature of quiescent prominences.

The exponents found here for the ranges above and below the break are consistent with strong and weak MHD turbulence, but, opposite to expectations, the exponents consistent with weak MHD turbulence are at smaller scales than those consistent with the strong MHD turbulence.
One hypothesis to explain this would be that the prevalence of flows found at the lengthscale associated with this change in exponent are key driving the change in turbulence regimes.
No great difference was found between the exponents of three separate regions of the prominence, which displayed different dynamical phenomena, or between the vertical and horizontal directions.

The turbulence in the prominence may be important for heating and diffusion processes.
The diffusion of the fluid across the magnetic field as a result of magnetic reconnection, reconnection diffusion, {is estimated to be $\eta_{\rm rec}\sim 4\times 10^{10}$\,cm$^2$\,s$^{-1}$ for appropriate parameters for a quiescent prominence}.
This is of similar order to the estimated ambipolar diffusion, and a few orders of magnitude greater than the Ohmic diffusion.
However, when estimating the heating rate as a result of the turbulence this was found to be small and as such unlikely to be of importance.}


\begin{acknowledgements}
The Authors would like to thank the anonymous referee for their invaluable comments.
Hinode is a Japanese mission developed and launched by ISAS/JAXA, with NAOJ as domestic partner and NASA and STFC (UK) as international partners. It is operated by these agencies in co-operation with ESA and NSC (Norway).
AH is supported by his STFC Ernest Rutherford Fellowship grant number ST/L00397X/1.
\end{acknowledgements}

\appendix
\section{Method for removing the stray light from the H$\alpha$ images}\label{Appen1}

Here we detail the techniques we used to process the H$\alpha$ intensity data before the creation of the Dopplergrams and transform it into velocities.
{The key point behind the} data processing applied here can be understood by looking at Equation \ref{dopgram_eqn}.
This equation is the difference of two intensities divided by {their} total.
Now imagine that both $I_+$ and $I_-$ are increased by some constants $A_+$ and $A_-$ as a result of {stray light (light reaching the camera pixels as a result of the optics)}.
Therefore, the nominator of Equation \ref{dopgram_eqn} increases by $A_+-A_-$ but the denominator increases by $A_++A_-$, which inherently reduces the value of the Dopplergram, ultimately reducing the velocity found.
Such a {stray light} component exists in the Hinode observations, so we created a model to remove it before making the Dopplergrams.

Figure \ref{background} panel (a) shows the level 1 data for H$\alpha$ -208m{\AA} with the intensity saturated to highlight the existence of {stray} light, as is obvious by the existence of a signal in the coronal region surrounding the prominence.
Panels (b) and (c) show the temporal fluctuations of the {stray} light intensity and the histogram of those fluctuations for the {pixels marked a and b} in panel (a).
The temporal fluctuations of the intensity can be broken down into three components, the approximately constant {stray} light modulated by the satellite motion of approximately $98$\,mins and the addition of a noise component, as well as an event where prominence material exists on that pixel.
The histogram of the intensity shown in panel (c), overplotted with a Gaussian distribution, has a standard deviation of $0.76$.

\begin{figure*}[ht]
\centering
\includegraphics[width=5.5cm]{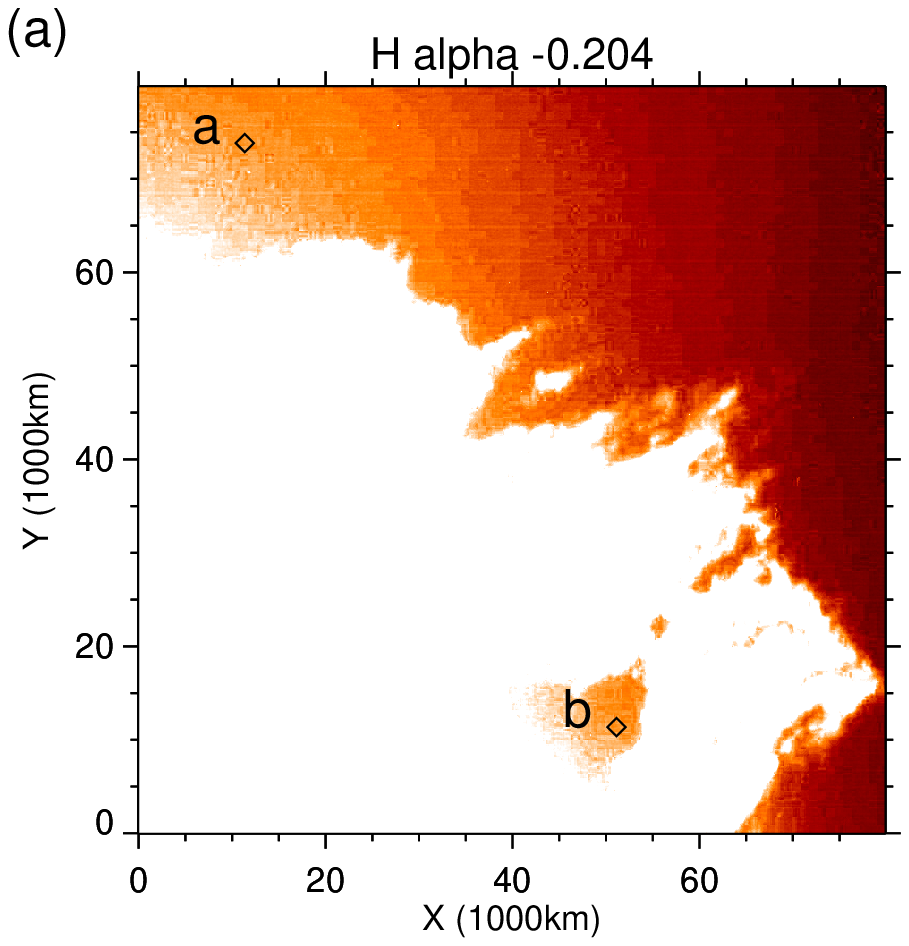}
\includegraphics[width=5.5cm]{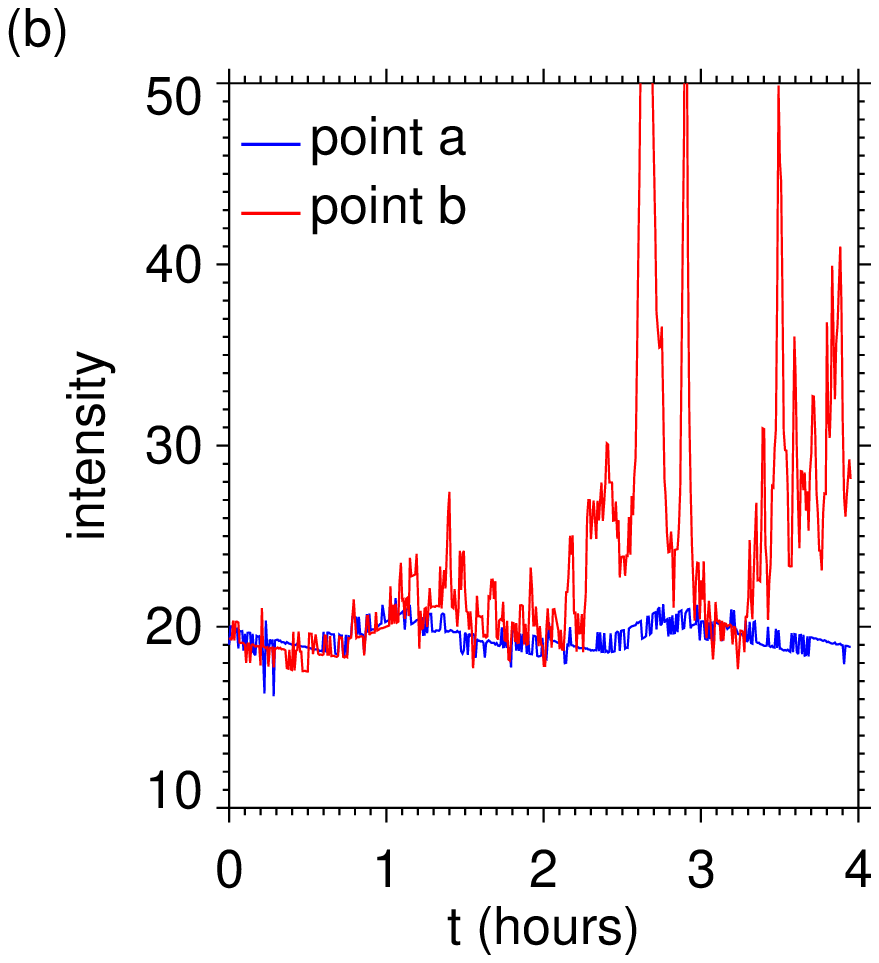}
\includegraphics[width=5.5cm]{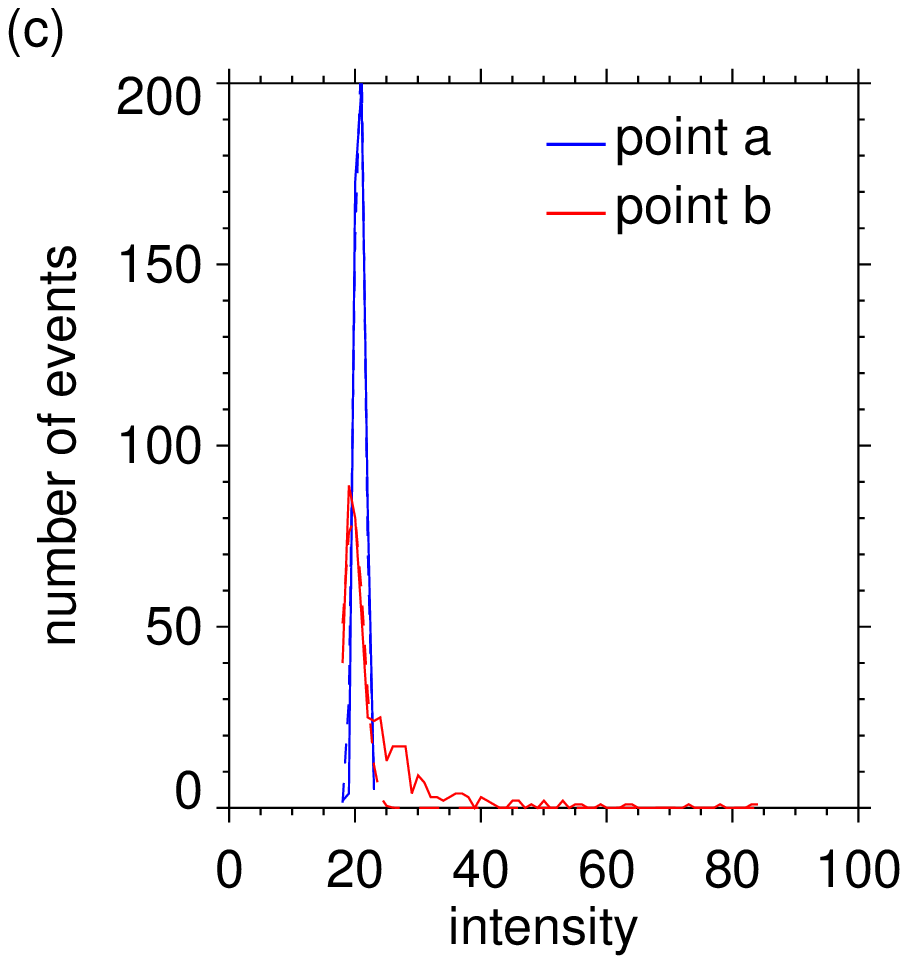}\\
\includegraphics[width=5.5cm]{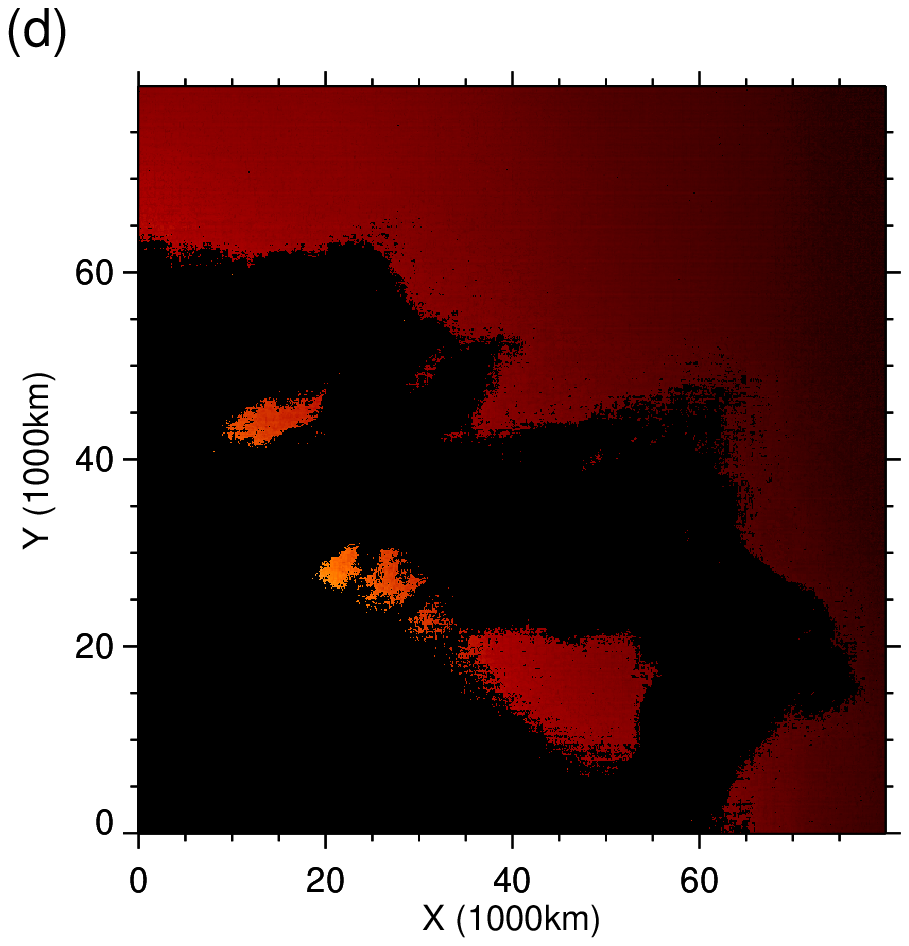}
\includegraphics[width=5.5cm]{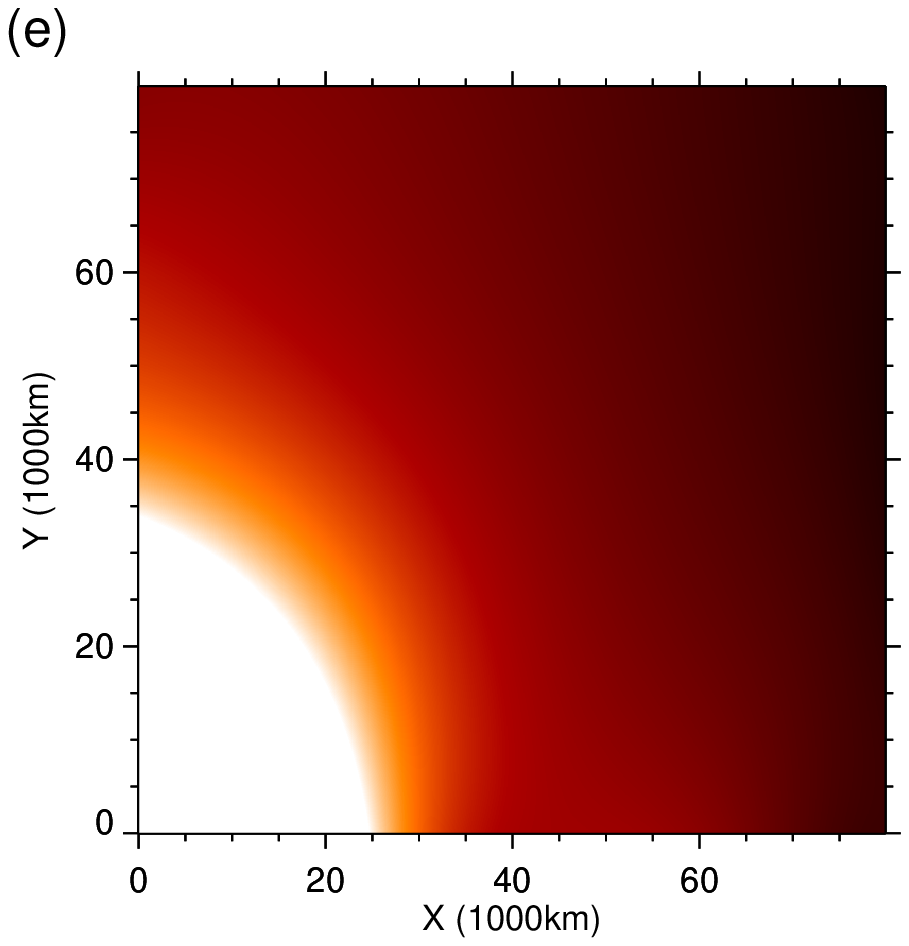}
\includegraphics[width=5.5cm]{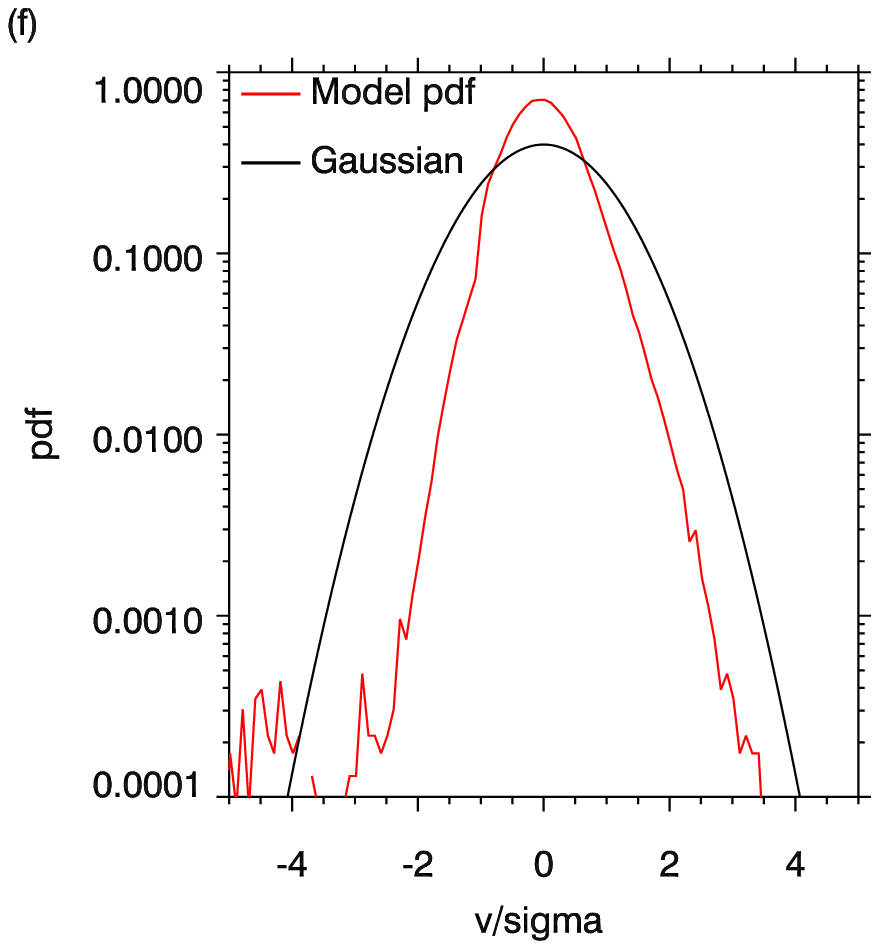}\\
\includegraphics[width=5.5cm]{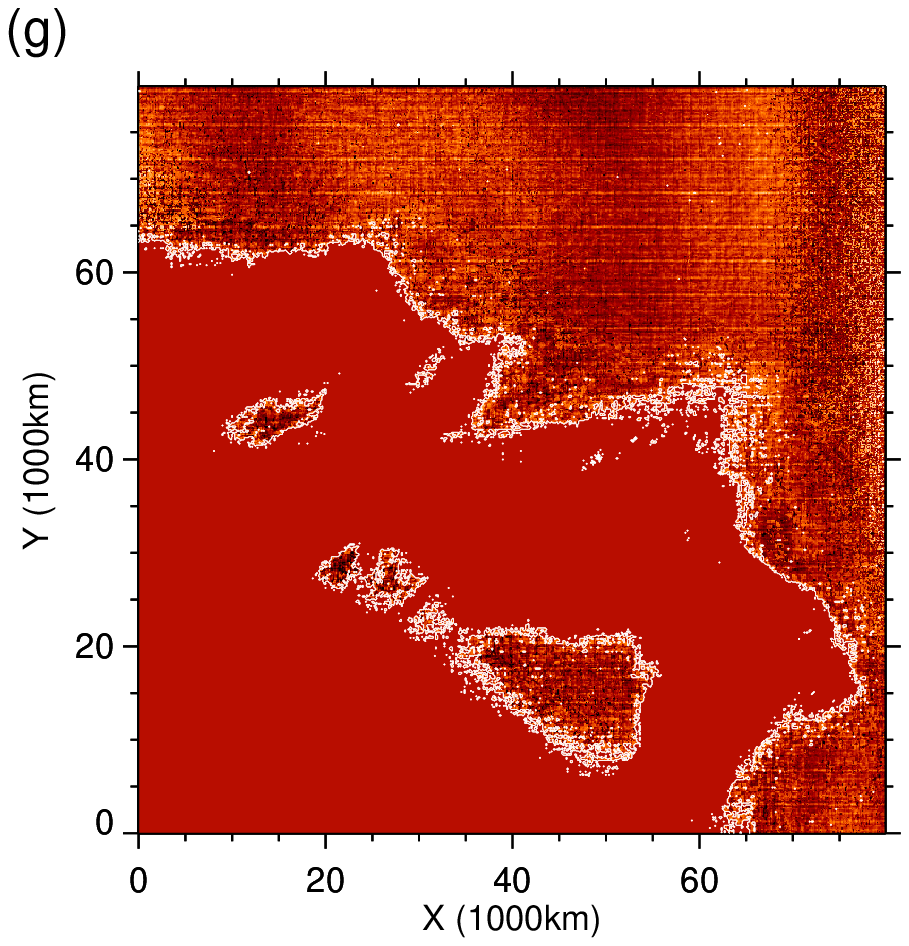}
\includegraphics[width=5.5cm]{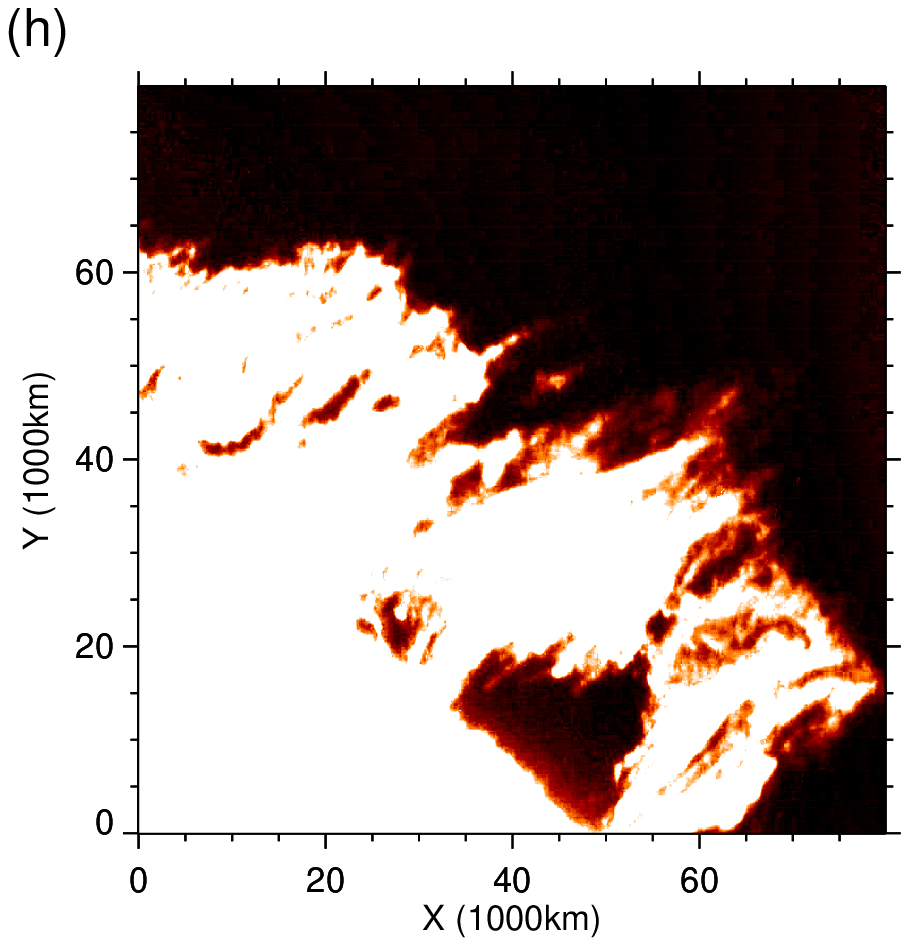}
\caption{(a) level 1 intensity (intensity range of $0$ to $25$), (b) temporal variation of intensity at points a and b, (c) {histograms of the intensities at points a and b}, (d) the mean background value (intensity range of $0$ to $25$), (e) the background model (intensity range of $0$ to $50$), (f) the histogram of the model values minus the mean background, (g) the 2D distribution of the difference between the model and mean background (colour range between -2 and 2 with white lines showing the boundary between regions with and without data) and (h) the level 1 intensity minus the background model (intensity range of $0$ to $25$).}
\label{background}
\end{figure*}

To model the {stray} light, it is necessary to identify the average value of the {stray} light in as many pixels as possible.
This is calculated by looking at the histogram for the intensity in each pixel and {identifying a Gaussian distribution} associated with the {stray} light in as many pixels as possible.
{Firstly, for each pixel, we impose the requirement for selection that there are more than forty exposures where the intensity of that pixel is less than 40.}
For these pixels, a histogram (of bins with width 1) is made and then a Gaussian fit is made to the seven bins associated with the smallest intensities.
If the centroid and width {of the fitted Gaussian distribution is} consistent with a well resolved peak in the intensity in this range, (e.g. see Fig. \ref{background} Panel c), then the value for the centroid of the Gaussian is taken as the real background intensity value and a 2D map can be made (see Fig. \ref{background} Panel d).
{Here we define a well resolved peak to mean that the half-width of the Gaussian to be less than $1.5$ and the position of the peak to be inside the second to sixth bin of fitting range under the condition that peak of the histogram across its whole range falls within the seven bins used for the fit.
For the red wing data shown in Fig. \ref{background} this initial process found the {stray} light value for 130912 pixels.}

Once all the {pixels whose stray light value satisfies these conditions} are determined, a two-dimensional fourth-order polynomial is fitted to the data, which gives an estimate for the {stray} light for each pixel.
However, as there may initially be a relatively small number of pixels for which we have managed to identify the stray light value in the prominence {(approximately $10^4$ pixels)}, the fit in this region may initially not be so good.
{To improve this, any pixels for which the stray light value has not been determined that have a minimum value plus $1\sigma$ (where the value of sigma is estimated from the noise fluctuations and found to be $\sigma =1$) smaller than the model value for that pixel, they have their minimum value plus $2\sigma$ set as the stray light value for the pixel.
Here the minimum value plus $2\sigma$ is taken because, through {trial} and error, it was found to find stray-light values that were consistent (within $\sim1\sigma$) with neighbouring pixels where the stray light had been determined through the initial method.}
The fit is repeated with these new values, and then the process described in the previous sentence is repeated.
It was found that repeating this iterative procedure ten times was sufficient to get approximate convergence for the model data {where the {stray} light value was determined for 229777 pixels in the red wing data with approximately $6\times 10^4$ pixels in the prominence region}.
{Figure \ref{background} panels (d) and (e) show the  data values for the pixels where an estimation of the stray light was obtained and the fitted model, respectively.}
Panels (f) and (g) give the 1D pdf and the 2D map of the residuals.
Both of these show that the error of the model can be taken as being at the noise level.
Panel (h) gives the H$\alpha$ $-208$\,m{\AA} intensity minus the {stray} light model value, from this it is clear the difference made to the intensity - and as a result the subsequent Dopplergrams - because this process has been followed.
It should be noted that this background intensity was found to be different for the $+$ and $-$ wings, requiring a separate model, following the same procedure, to be made for each.

\section{Validation of line width}\label{line_width}

For this study, to produce the velocity proxy {using Equation \ref{vel_trans}}, it was necessary to assume a line width.
We selected a line width $\sigma_{\lambda}$ based on the thermal velocity of hydrogen at $8000$\,K.
However, it is necessary to determine what, if any, effect this assumption has on the results.
To this end, we performed a series of calculations using different values for the line width and compare the second order structure function for the velocity increments ($\delta_r v$) across the whole prominence that are produced.

Along with {velocity increments calculated from the velocity distribution} given by a {line width with thermal velocity with} temperature of $8000$\,K, for this part of the investigation we also use temperatures of $6000$\,K and $10^4$\,K, and a distribution obtained when the temperature of the prominence plasma varies with a normal distribution centred on $8000$\,K with a standard deviation of $1000$\,K {denoted G$_{10^3}8000$\,K}.
{Here the source of the line width as been assumed to be a thermal velocity, but in reality it is likely that superposition of non-thermal motions along the line-of-sight will also lead to {line broadening} and will play a role giving the Doppler width $v_{\rm DW}$ of the line as $v_{\rm DW}^2=v_{\rm T}^2+v_{\rm los}^2$ where $v_{\rm T}$ is the thermal velocity and $v_{\rm los}$ is broadening by the line-of-sight motions.
Therefore, when we talk of a line {width} based on the thermal velocity of $8000$\,K this is equivalent to saying that the velocity associated with the Doppler width is equal to that of the thermal velocity of a fluid at $8000$\,K, that is $v_{\rm DW}=v_{8000\rm K}$.
The purpose of this appendix is to understand how different line widths, whatever their cause, could result in changes in the results of this paper.}
Figure \ref{width_check} gives the {second order structure functions for all of these velocity increment distributions.}
{All of these structure functions are calculated from a single snapshot from the data.}

\begin{figure}[ht]
\centering
\includegraphics[width=8cm]{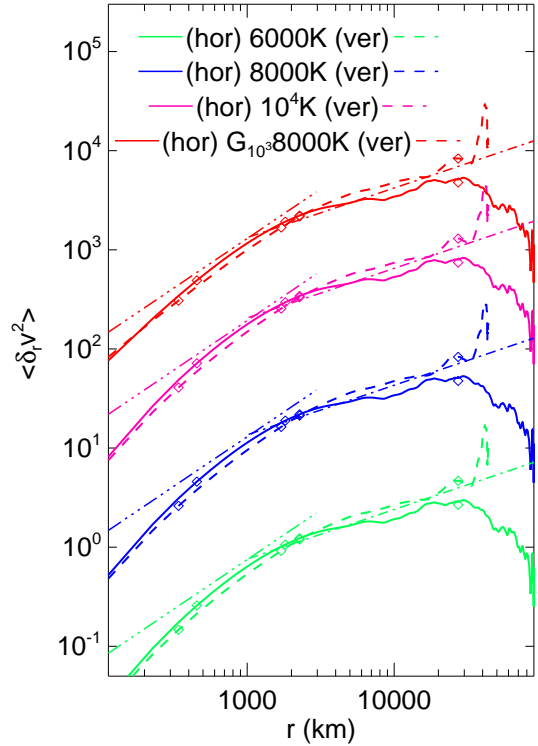}
\caption{Plot of $\langle\delta_r v^2\rangle$ for the four different models of the line width. From bottom to top these are the $6000$\,K line width (green), the $8000$\,K line width (blue), the $10^4$\,K line width (pink) and the randomly distributed temperature model (red). The solid line shows {the structure function calculated from the horizontal velocity increments and the dashed line the structure function calculated from the vertical velocity increments}. It should be noted that that other than the $6000$\,K distribution, the others have been rescaled to allow for them to be plotted clearly. To highlight the similarity in the scalings of the four distributions, two power laws have been overplotted: $r^{1/2}$ in the dot-dash line and $r^{1}$ in the triple dot-dash line. {The triangles and diamonds show the range over which respectively Exponent 1 and Exponent 2, as shown in Table \ref{table2}, are calculated.}}
\label{width_check}
\end{figure}

As can be seen in Fig. \ref{width_check}, the distributions of the second order structure functions do not present any large changes as a result of the different assumptions of line width.
Especially important is the fact that the general exponents presented by the distributions, both at the smaller and large spatial scales, are consistent.
{The exponents calculated from these structure functions between the ranges given by the triangles and diamonds are listed in Table \ref{table2}.
Exponent 1 is the value below the break and Exponent 2 is the value above.
The exponent values do not show any variation to a level that would change the interpretation of the results in this paper.
The value of $\langle \delta_r v^2 \rangle$ at r=3000\,km is also given to shown how the magnitude changes as a result of the line-width assumption, so, for example, the heating estimated in Section \ref{heat_sec} is likely to have an uncertainty of a factor $3$ to $5$ due to the uncertainty of the line-width.}
Therefore, we can be confident that the results we present are not dependent of our choice of line width.

\begin{table*}
\begin{center}
\caption{{Exponents of the power laws found for $\langle \delta_r v^2 \rangle$ and the magnitude at $r=3000$\,km for the four different line widths. Values for both the horizontal (left) and vertical (right) structure functions are given.}}
\begin{tabular}{ c  c  c  c  c }
\hline
$\sigma_{\lambda}$ & $6000$\,K & $8000$\,K & $10^4$\,K & G$_{10^3}8000$\,K\\ \hline
Exponent 1  & $1.14$ / $1.28$ & $1.14$ / $1.28$ & $1.14$ / $1.28$ & $1.10$ / $1.20$   \\
Exponent 2  & $0.37$ / $0.49$ & $0.37$ / $0.49$ & $0.37$ / $0.49$  & $0.37$ / $0.49$   \\
$\langle \delta_r v^2 \rangle$ at r=3000\,km  & $2.0$ / $1.5$  (km/s)$^2$& $3.6$ / $2.6$ (km/s)$^2$ & $5.62$ / $4.1$ (km/s)$^2$& $3.6$ / $2.6$ (km/s)$^2$ \\
\hline 
\end{tabular}
 
\label{table2}
\end{center}
\end{table*}

\end{document}